\documentclass[reprint,aps,prb,superscriptaddress]{revtex4-1}
\usepackage[unicode=true, breaklinks=false, pdfborder={0 0 1}, backref=false, colorlinks=true, linkcolor=blue, urlcolor=blue, citecolor=blue]{hyperref}
\usepackage{amsmath}
\usepackage{amssymb}
\usepackage{natbib}
\usepackage{graphicx}
\usepackage{color}
\usepackage{colordvi}
\usepackage{amsopn}
\usepackage[utf8x]{inputenc}

\newcommand{\func}[1]{\operatorname{#1}}
\setcitestyle{numbers,square}
\begin{document}

\title{Chiral coupling of magnons in waveguides}
\author{Tao Yu}
\affiliation{Kavli Institute of Nanoscience, Delft University of Technology, 2628 CJ Delft, The Netherlands}
\author{Xiang Zhang}
\affiliation{Kavli Institute of Nanoscience, Delft University of Technology, 2628 CJ Delft, The Netherlands}
\author{Sanchar Sharma}
\affiliation{Kavli Institute of Nanoscience, Delft University of Technology, 2628 CJ Delft, The Netherlands}
\author{Yaroslav M. Blanter}
\affiliation{Kavli Institute of Nanoscience, Delft University of Technology, 2628 CJ Delft, The Netherlands}
\author{Gerrit E. W. Bauer}
\affiliation{Institute for Materials Research $\&$ WPI-AIMR $\&$ CSRN,
	Tohoku University, Sendai 980-8577, Japan}
\affiliation{Kavli Institute of Nanoscience, Delft University of Technology, 2628 CJ Delft, The Netherlands}
\date{\today }

\begin{abstract}
We theoretically investigate the collective excitation of multiple
(sub)millimeter-sized ferromagnets mediated by waveguide photons. By the
position of the magnets in the waveguide, the magnon-photon coupling can be
tuned to be chiral, i.e., magnons only couple with photons propagating in
one direction, leading to asymmetric transfer of angular momentum and energy
between the magnets. A large imbalance in the magnon number distribution
over the magnets can be achieved with a long chain of magnets, which
concentrate at one edge. The chain also supports standing waves with low
radiation efficiency that is inert to the chirality.
\end{abstract}

\maketitle

\affiliation{Kavli Institute of Nanoscience, Delft University of Technology,
2628 CJ Delft, The Netherlands}

\affiliation{Kavli Institute of Nanoscience, Delft University of Technology,
2628 CJ Delft, The Netherlands}

\affiliation{Kavli Institute of Nanoscience, Delft University of Technology,
2628 CJ Delft, The Netherlands}

\affiliation{Kavli Institute of Nanoscience, Delft University of Technology,
2628 CJ Delft, The Netherlands}

\affiliation{Institute for Materials Research $\&$ WPI-AIMR $\&$ CSRN,
	Tohoku University, Sendai 980-8577, Japan} 
\affiliation{Kavli Institute of
Nanoscience, Delft University of Technology, 2628 CJ Delft, The Netherlands}


\section{Introduction}

Magnetic insulators are promising materials for low-dissipation information
technology with magnons, the elementary excitation of magnetic order, rather
than electrons \cite{magnonics1,magnonics2,magnonics3,magnonics4}. The long
lifetime of magnons in high-quality magnetic insulators such as yttrium iron
garnet (YIG) \cite{low_damping} are suitable for data storage, logic, and
medium-distance interconnects, but cannot compete with photons in terms of
speed and coherence lengths. Coupled magnon-photon systems are therefore
promising for quantum communication over large distances \cite%
{quantum_magnon}. The interface to conventional electronics are metal
contacts that allow magnons to interact with conduction electrons by
interfacial exchange interaction, giving rise to spin pumping and spin
transfer torques \cite{spin_pumping_electron,non_local}. Magnons in separate
nano-magnets couple by the long-range dipolar interaction, which gives rise
to novel phenomena \cite{Yu1,Yu2,Yu3}.

Strong coherent coupling between photons in high-quality cavities and spin
ensembles such as NV centers in diamond \cite{Kubo_NV,Amsuss_NV}, rare earth
ions \cite{Probst_REI,Schuster_REI}, and ferromagnets \cite%
{first_2010,second,third,fourth}, is attractive because of its potential for
quantum memories \cite{MagnonDarkModes} and transducers. While a (nearly)
closed cavity can have very long photon lifetimes, efficient photon
transport requires an open waveguide, which is the main object of the present study.
Coherent microwave emission from a precessing magnetization of a ferromagnet
in a waveguide can be measured via the additional damping of magnons \cite%
{RD1,RD2A,RD2,RD3,Magnon_radiation} on top of the intrinsic Gilbert damping.
The Larmor precession of the magnetization couples preferentially to photons
with the same polarization. Due to the tunable ellipticity of the AC
magnetic field, magnets at certain locations in a waveguide (to be discussed
in the main text) also couple preferentially to photons propagating in one
direction. Such a chiral coupling \cite{Jackson} of atoms and quantum dots
with optical photons attracts much attention \cite%
{chiral_optics1,chiral_optics2,chiral_emitter,chiral_optics3,chiral_optics4,chiral_optics5,chiral_review}%
.

\begin{figure}[ht]
{\includegraphics[width=\columnwidth]{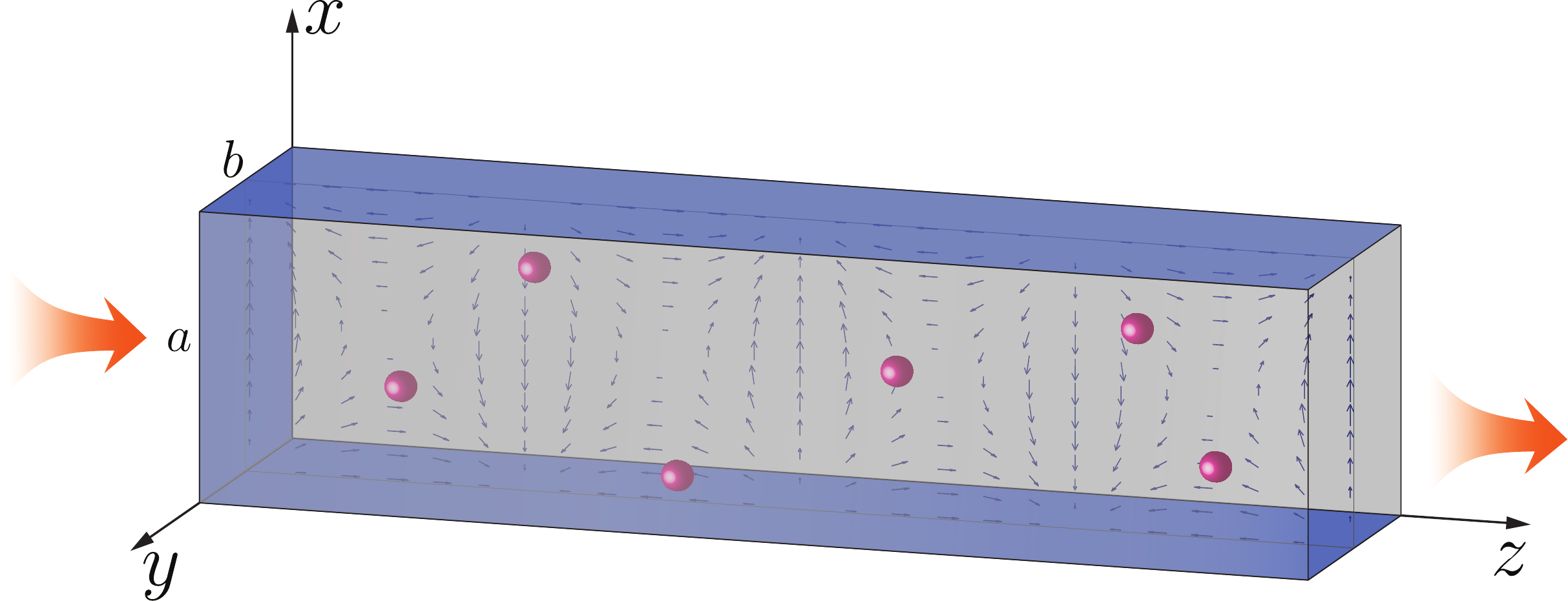}}
\caption{An ensemble of magnets in a waveguide along the $\mathbf{z}$%
-direction. The input photon shown by the red arrow experiences scattering
by magnets, and its transmission can be used to detect the magnon dynamics.}
\label{fig:setup}
\end{figure}

Here we study a collection of magnetic particles placed in a microwave
waveguide \cite{level_attraction1,Magnon_radiation,Yao_Yu}, as shown in Fig.~%
\ref{fig:setup}. The radiation emitted by a magnet drives typically all the
other magnets, leading to an effective long-range dissipative coupling,
reminiscent of but very different from the coherent coupling in a closed
cavity \cite{MagnonDarkModes,Babak}. The coupling mediated by travelling
photons in atomic ensembles \cite%
{subradiance1,subradiance2,subradiance3,subradiance4,subradiance5,subradiance6}
causes collective super- and sub-radiance. Here, we discuss analogous modes
in macroscopic magnonic systems but incorporating the chirality, which
can be probed by microwaves at room temperature.

We show that magnets can couple chirally to waveguide photons, leading to nonreciprocal
magnon-magnon interaction \cite{chiral_review}. For given locations in a
waveguide, one magnet can affect another one without back-action \cite%
{chiral_emitter}. We predict an imbalance of the magnon population in two
spheres of up to one order of magnitude, which can be significantly enhanced
in a chain of magnets. We study the collective excitations of up to $\sim
100 $ magnets, focusing on super-radiant and sub-radiant modes, i.e.
modes with very high or low radiation efficiency. We find that the
superradiant states \cite%
{subradiance1,subradiance2,subradiance3,subradiance4,subradiance5} are well
localized at the edge of the chain \cite{subradiance1,non_hermitian5}. In
contrast, the lowest subradiant states are standing-wave--like and centered
in the chain and are only weakly affected by the chirality of the coupling.
In the accompanying Letter \cite{PRL_submission}, we introduce this effect
and focus on the new functionality of generating very large magnon
amplitudes at the edges of magnetic chains in a waveguide and work out its
enhancement by chirality. Here we formulate the theory and observables for the physical
properties of the collective modes for one, two and many spheres in a
waveguide, such as the microwave transmission spectra.

This paper is organized as follows. We introduce the model in Sec.~\ref%
{general} including the Hamiltonian and photon scattering matrix for a
general waveguide geometry and positions of the magnets. After a focus on
magnon-photon coupling in a rectangular waveguide in Sec.~\ref{Sec:Rect}, we
address the radiative damping of magnets in Sec.~\ref{one_sphere}. In Sec.~%
\ref{two_sphere}, we discuss the transmission of a waveguide with two
magnetic spheres, introducing the concept of imbalanced pumping. We derive
collective modes with super- and sub-radiance in long magnetic chains in
Sec.~\ref{many_sphere}. Finally, Sec.~\ref{summary} contains a discussion of
the results and conclusions.

\section{Model}

\label{general}

We focus here on magnets that are small enough compared with the photon
wavelength such that only the homogeneous collective excitation or Kittel
mode couples with the microwave photon \cite{Kittel_Orig,Kittel_book}. We
consider a waveguide infinite in the $\mathbf{z}$-direction with a
rectangular cross-section from $(0,0)$ to $(a,b)$, as shown in Fig.~\ref%
{fig:setup}. We assume metallic boundaries, i.e. the electric field parallel
to the surface vanishes. There are $N$ equivalent magnets with gyromagnetic
ratio $-\tilde{\gamma}$, saturation magnetization $M_{s}$, and volume $V_{s}$%
. Their centers are at $\mathbf{r}_{i}=(\pmb{\rho}_{i},z_{i})$, where $%
\pmb{\rho}=(x,y)$ is the position in the waveguide's cross-section.

The dynamics is governed by the Hamiltonian $\hat{H}=\hat{H}_{\mathrm{em}}+%
\hat{H}_{\mathrm{m}}+\hat{H}_{\mathrm{int}}$, with electromagnetic
contribution 
\begin{equation}
\hat{H}_{\mathrm{em}}=\int \left[ \frac{\epsilon _{0}}{2}\mathbf{E}(\mathbf{r%
})\cdot \mathbf{E}(\mathbf{r})+\frac{\mu _{0}}{2}\mathbf{H}(\mathbf{r})\cdot 
\mathbf{H}(\mathbf{r})\right] d\mathbf{r},  \label{CHam:em}
\end{equation}%
the magnetic part 
\begin{equation}
\hat{H}_{\mathrm{m}}=-\mu _{0}\int \left[ H_{\mathrm{app}}(\mathbf{r})M_{y}(%
\mathbf{r})+\mathbf{H}_{\mathrm{eff}}(\mathbf{r})\cdot \mathbf{M}(\mathbf{r})%
\right] d\mathbf{r},  \label{CHam:m}
\end{equation}%
and the magnon-photon interaction 
\begin{equation}
\hat{H}_{\mathrm{int}}=-\mu _{0}\int \mathbf{H}(\mathbf{r})\cdot \mathbf{M}(%
\mathbf{r})d\mathbf{r}.  \label{CHam:int}
\end{equation}%
The time-dependence is implicit. Here, $\{\mathbf{E},\mathbf{H}\}$ represent
the electric and magnetic fields of the photons in the waveguide, $\mathbf{H}%
_{\mathrm{eff}}(\mathbf{r})$ is the sum of dipolar and exchange interaction 
\cite{Landau}, $\mathbf{M}$ is the magnetization, $\epsilon _{0}$ and $\mu
_{0}$ are the permittivity and permeability of the free space, and $H_{%
\mathrm{app}}(\mathbf{r})\mathbf{y}$ denotes the static applied field that
saturates the magnetizations.

The Hamiltonian gives the Maxwell equations \cite{Jackson}, 
\begin{align}
\boldsymbol{\nabla }\times \mathbf{E}& =-\mu _{0}\frac{\partial \mathbf{H}}{%
\partial t},\ ~\boldsymbol{\nabla }\times \mathbf{H}=\epsilon _{0}\frac{%
\partial \mathbf{E}}{\partial t},  \notag \\
\boldsymbol{\nabla }\cdot \mathbf{H}& =-\boldsymbol{\nabla }\cdot \mathbf{M}%
,\ ~\boldsymbol{\nabla }\cdot \mathbf{E}=0,
\end{align}%
and the Landau-Lifshitz (LL) equation \cite{Landau} 
\begin{equation}
\frac{\partial \mathbf{M}}{\partial t}=-\tilde{\gamma}\mu _{0}\mathbf{M}%
\times \left( \mathbf{H}+\mathbf{H}_{\mathrm{eff}}+H_{\mathrm{app}}\mathbf{y}%
\right) .
\end{equation}

The electromagnetic fields can be expanded in photon operators, 
\begin{equation}
\mathbf{H}(\mathbf{r})=\sum_{\lambda }\int_{-\infty }^{\infty }\left( %
\pmb{\cal H}_{k}^{\lambda }(\pmb{\rho})e^{ikz}\hat{p}_{k}^{\lambda }+\mathrm{%
h.c.}\right) \frac{dk}{\sqrt{2\pi }},  \label{HExp}
\end{equation}%
with $\pmb{\cal H}_{k}^{\lambda }(\pmb{\rho})$ being the eigenmodes for the
magnetic field in the waveguide \cite{Jackson}, and similarly for the
electric field with $\mathbf{H}\rightarrow \mathbf{E}$ and $\pmb{\cal H}%
\rightarrow \pmb{\cal E}$. Here $k$ denotes the momentum in the $\mathbf{z}$%
-direction, and $\lambda $ represents the mode structure (including the
polarization). The photon operators satisfy the field commutation relations 
\begin{equation}
\left[ \hat{p}_{k}^{\lambda },\hat{p}_{k^{\prime }}^{\lambda ^{\prime
}\dagger }\right] =\delta (k-k^{\prime })\delta _{\lambda \lambda ^{\prime
}}.
\end{equation}%
The Cartesian components of the eigenmodes $\pmb{\cal H}_{k}^{\lambda }(%
\pmb{\rho})$ and $\pmb{\cal E}_{k}^{\lambda }(\pmb{\rho})$ in a 
waveguide satisfy the orthonormality relations \cite{Jackson}, 
\begin{align}
\int \left( {\mbox{\boldmath$\cal H$\unboldmath}}_{k,x}^{\lambda \ast }{%
\mbox{\boldmath$\cal H$\unboldmath}}_{k,x}^{\lambda ^{\prime }}+{%
\mbox{\boldmath$\cal H$\unboldmath}}_{k,y}^{\lambda \ast }{%
\mbox{\boldmath$\cal H$\unboldmath}}_{k,y}^{\lambda ^{\prime }}\right) d%
\pmb{\rho}& =\frac{A_{k}^{\lambda }}{(Z_{k}^{\lambda })^{2}}\delta _{\lambda
\lambda ^{\prime }},  \notag \\
\int \mathcal{H}_{k,z}^{\lambda \ast }\mathcal{H}_{k,z}^{\lambda ^{\prime }}d%
\pmb{\rho}& =\frac{\gamma _{\lambda }^{2}A_{k}^{\lambda }}{%
k^{2}(Z_{k}^{\lambda })^{2}}\delta _{\lambda \lambda ^{\prime }},\hspace{%
0.4cm}(\mathrm{TE)}  \notag \\
\int \left( {\mbox{\boldmath$\cal E$\unboldmath}}_{k,x}^{\lambda \ast }{%
\mbox{\boldmath$\cal E$\unboldmath}}_{k,x}^{\lambda ^{\prime }}+{%
\mbox{\boldmath$\cal E$\unboldmath}}_{k,y}^{\lambda \ast }{%
\mbox{\boldmath$\cal E$\unboldmath}}_{k,y}^{\lambda ^{\prime }}\right) d%
\pmb{\rho}& =A_{k}^{\lambda }\delta _{\lambda \lambda ^{\prime }},  \notag \\
\int \mathcal{E}_{k,z}^{\lambda \ast }\mathcal{E}_{k,z}^{\lambda ^{\prime }}d%
\pmb{\rho}& =\frac{\gamma _{\lambda }^{2}A_{k}^{\lambda }}{k^{2}}\delta
_{\lambda \lambda ^{\prime }},\hspace{0.8cm}(\mathrm{TM).}
\label{orthogonal}
\end{align}%
Here, $Z_{k}^{\lambda }=\mu _{0}\Omega _{k}^{\lambda }/k$ and $k/(\epsilon
_{0}\Omega _{k}^{\lambda })$ are, respectively, the impedances for the TE
and TM modes \cite{Jackson}, $A_{k}^{\lambda }=\hbar \Omega _{k}^{\lambda
}/(2\epsilon _{0})$ and $\hbar /(2\epsilon _{0}\Omega _{k}^{\lambda })$ for
the TE and TM modes with $\Omega _{k}^{\lambda }$ being the eigen frequency,
and 
\begin{equation}
\gamma _{\lambda }^{2}=(\Omega _{k}^{\lambda })^{2}/c^{2}-k^{2}.
\label{spectra}
\end{equation}%
TE (TM), i.e. transverse electric (magnetic) polarization, refers to the
case when the electric (magnetic) field is perpendicular to the $\mathbf{z}$%
-direction. It is noted that these normalizations are chosen such that the
Hamiltonian Eq.~(\ref{CHam:em}) satisfies (up to a constant) 
\begin{equation}
\hat{H}_{\mathrm{em}}=\sum_{\lambda }\int \hbar \Omega _{k}^{\lambda }\hat{p}%
_{k}^{\lambda \dagger }\hat{p}_{k}^{\lambda }dk.  \label{Ham:em}
\end{equation}%
We assume the losses in high-quality waveguide to be small compared to the
magnetic dissipation and not important on the length scale of interest.

The magnetization $\mathbf{M}(\mathbf{r})$ is confined to the magnets that
are much smaller than typical photon wavelengths and waveguide dimensions
(usually $>1$~cm), such that the magnetic field is a constant inside each
magnet. The excitations of the (linearized) magnetic Hamiltonian are spin
waves, or its quanta, magnons. For magnets with axial symmetry around the
magnetization, the microwaves couple strongly only with the Kittel mode,
i.e. the uniform precession of the magnetization and we disregard other
modes in the following. We quantize the magnetization as \cite%
{squeezed_magnon,HP,surface_roughness} 
\begin{align}
M_{j,z}-iM_{j,x}& =\sqrt{\frac{2\hbar \tilde{\gamma}M_{s}}{V_{s}}}\hat{m}%
_{j},  \notag \\
M_{j,y}& =M_{s}-\frac{\hbar \tilde{\gamma}}{V_{s}}\hat{m}_{j}^{\dagger }\hat{%
m}_{j},
\label{HP}
\end{align}%
where $\hat{m}_{j}$ is the annihilation operator for a Kittel magnon in the $%
j$-th magnet with $j\in \{1,\dots ,N\}$. The coefficients are chosen to
ensure that $\mathbf{M}_{j}\cdot \mathbf{M}_{j}\approx M_{s}^{2}$ and the
magnetic Hamiltonian Eq. (\ref{CHam:m}), up to a constant due to zero-point
fluctuations, becomes 
\begin{equation}
\hat{H}_{\mathrm{m}}=\sum_{j=1}^{N}\hbar \omega _{j}\hat{m}_{j}^{\dagger }%
\hat{m}_{j},  \label{Ham:m}
\end{equation}%
where $\omega _{j}=\tilde{\gamma}\mu _{0}\left[ H_{\mathrm{app}}(\mathbf{r}%
_{j})+H_{\mathrm{eff}}(\mathbf{r}_{j})\right] $ with $H_{\mathrm{eff}%
}=N_{y}H_{\mathrm{app}}$ for axially symmetric magnets ($N_{y}$ is the
demagnetization factor).

Inserting Eqs.~(\ref{HExp}) and (\ref{HP}) into the interaction
Hamiltonian Eq. (\ref{CHam:int}), 
\begin{equation}
\hat{H}_{\mathrm{int}}=\sum_{j\lambda }\int \left[ \hbar g_{j}^{\lambda }(k)%
\hat{p}_{k}^{\lambda }\hat{m}_{j}^{\dagger }+\mathrm{h.c.}\right] \frac{dk}{%
\sqrt{2\pi }},
\end{equation}%
with coupling constant 
\begin{equation}
g_{j}^{\lambda }(k)=\mu _{0}\sqrt{\frac{\tilde{\gamma}M_{s}V_{s}}{2\hbar }}%
e^{ikz_{j}}\left[ i\mathcal{H}_{k,x}^{\lambda }(\pmb{\rho}_{j})-\mathcal{H}%
_{k,z}^{\lambda }(\pmb{\rho}_{j})\right] .  \label{coupling}
\end{equation}%
The distributed magnets experience different phases when their distance is
not much smaller than the photon wavelength. We can tune coupling strength
and chirality by the position of the magnets $\pmb{\rho}_{j}$, see Sec. \ref%
{Sec:Rect}.

\subsection{Equations of motion}

\label{Sec:EOMs}

From the Hamiltonian $\hat{H}=\hat{H}_{\mathrm{em}}+\hat{H}_{\mathrm{m}}+%
\hat{H}_{\mathrm{int}}$, we obtain the equation of motion for photons by the
Heisenberg equation 
\begin{equation}
\frac{d\hat{p}_{k}^{\lambda }}{dt}=-i\Omega _{k}^{\lambda }\hat{p}%
_{k}^{\lambda }-i\sum_{j}\frac{g_{j}^{\lambda \ast }(k)}{\sqrt{2\pi }}\hat{m}%
_{j}.
\end{equation}%
The solutions are 
\begin{equation}
\hat{p}_{k}^{\lambda }(t)=\hat{p}_{k,\mathrm{in}}^{\lambda }e^{-i\Omega
_{k}^{\lambda }t}-\sum_{j}\frac{ig_{j}^{\lambda \ast }(k)}{\sqrt{2\pi }}%
\int_{-\infty }^{t}\hat{m}_{j}(\tau )e^{-i\Omega _{k}^{\lambda }(t-\tau
)}d\tau ,  \label{RetAp}
\end{equation}%
where $\hat{p}_{k}(-\infty )\equiv \hat{p}_{k,\mathrm{in}}^{\lambda }$ is
the microwave input \cite{input_output1,input_output2}. The first term is
the free evolution and the second term is the (spontaneous and stimulated)
radiation generated by magnons. The output field $\hat{p}_{k,\mathrm{out}%
}^{\lambda }=\lim_{t\rightarrow \infty }\hat{p}_{k}^{\lambda }(t)e^{i\Omega
_{k}^{\lambda }t}$ then reads 
\begin{equation}
\hat{p}_{k,\mathrm{out}}^{\lambda }=\hat{p}_{k,\mathrm{in}}^{\lambda
}-i\sum_{j}\frac{g_{j}^{\lambda \ast }(k)}{\sqrt{2\pi }}\int_{-\infty
}^{\infty }d\tau \ \hat{m}_{j}(\tau )e^{i\Omega _{k}^{\lambda }\tau }.
\label{IO:Basic}
\end{equation}

The magnon dynamics is governed by equation of motion 
\begin{equation}
\frac{d\hat{m}_{j}}{dt}=-i\omega _{j}\hat{m}_{j}-\hat{\mathcal{D}}_{\mathrm{%
int},j}-\hat{\mathcal{D}}_{\mathrm{ph},j},  \label{LLG}
\end{equation}%
where 
\begin{align}
\hat{\mathcal{D}}_{\mathrm{int},j}& =\frac{\kappa _{j}}{2}\hat{m}_{j}+\sqrt{%
\kappa _{j}}\hat{N}_{j},  \label{IntDamp} \\
\hat{\mathcal{D}}_{\mathrm{ph},j}& =i\sum_{\lambda }\int \frac{dk}{\sqrt{%
2\pi }}g_{j}^{\lambda }(k)\hat{p}_{k}^{\lambda },  \label{ExtDamp}
\end{align}%
equivalent to the linearized Landau-Lifshitz-Gilbert (LLG) equation. Here
the linewidth $\kappa _{j}=2\alpha _{G}\omega _{j},$ where $\alpha _{G}$ is
the Gilbert damping parameter. Each magnet $j$ is connected to an intrinsic
bath of phonons and other magnons, which generates the thermal torque $\hat{%
\mathcal{D}}_{\mathrm{int},j}$. We model this interaction by a Markovian
processes with intrinsic linewidth $\kappa _{j}$ and white noise $\hat{N}%
_{j} $ satisfying $\left\langle \hat{N}_{j}\right\rangle =0$, $\left\langle 
\hat{N}_{j}^{\dagger }(t)\hat{N}_{j}(t^{\prime })\right\rangle =n_{j}\delta
(t-t^{\prime })$ and $\left\langle \hat{N}_{j}(t)\hat{N}_{j}^{\dagger
}(t^{\prime })\right\rangle =(n_{j}+1)\delta (t-t^{\prime }),$ where 
\begin{equation}
n_{j}=\left[ \exp \left( \frac{\hbar \omega _{j}}{k_{B}T}\right) -1\right]
^{-1}
\end{equation}%
is the thermal occupation of magnons at a global temperature $T$. In the
absence of coupling between different magnets by a waveguide, $\hat{\mathcal{%
D}}_{\mathrm{ph},j}=0$ and all magnons are Gibbs distributed at equilibrium 
\cite{input_output1}.

When magnons are coupled by photons, the torque $\hat{\mathcal{D}}_{\mathrm{%
ph},j}$ can be split as 
\begin{equation}
\hat{\mathcal{D}}_{\mathrm{ph},j}(t)=\hat{T}_{j}(t)+i\sum_{l}\int_{-\infty
}^{t}d\tau \ \tilde{\Sigma}_{jl}(t-\tau )\hat{m}_{l}(\tau ),
\end{equation}%
where the first term is generated by the photon input, 
\begin{equation}
\hat{T}_{j}(t)=i\sum_{\lambda }\int \frac{dk}{\sqrt{2\pi }}g_{j}^{\lambda
}(k)\hat{p}_{k,\mathrm{in}}^{\lambda }e^{-i\Omega _{k}^{\lambda }t},
\end{equation}%
while the second term describes the photon-mediated coupling 
\begin{equation}
\tilde{\Sigma}_{jl}(t-\tau )=-i\sum_{\lambda }\int \frac{dk}{2\pi }%
g_{j}^{\lambda }(k)g_{l}^{\lambda \ast }(k)e^{-i\Omega _{k}^{\lambda
}(t-\tau )},
\end{equation}%
which can be interpreted as (real or virtual) $(\lambda ,k)$-mode photon
emission from magnet $l$ with amplitude $g_{l}^{\lambda \ast }(k)$ followed
by absorption in magnet $j$ with amplitude $g_{j}^{\lambda }(k).$ The
interaction is retarded by the finite light velocity. However, even for
large distances $r_{jl}<1$~m, $\kappa _{j}r_{jl}/c<0.02$, where $\kappa
_{j}=2\pi \times 1\,$MHz is a typical magnon linewidth, so $\tilde{\Sigma}%
_{jl}(t-\tau )$ decays much faster than the magnon envelope dynamics. For
short times $|t-\tau |<r_{jl}/c$ the magnons may be assumed to move
coherently $\hat{m}_{l}(\tau )\approx \hat{m}_{l}(t)e^{i\omega _{l}(t-\tau
)} $. This adiabatic approximations simplifies Eq. (\ref{LLG}) to%
\begin{equation}
\frac{d\mathcal{\hat{M}}}{dt}=-i\tilde{\omega}\hat{\mathcal{M}}-i\Sigma \hat{%
\mathcal{M}}-\hat{\mathcal{T}}-\hat{\mathcal{N}},  \label{EOM:Mags}
\end{equation}%
introducing the column vectors for magnetization $\mathcal{\hat{M}}=\left( 
\hat{m}_{1},\dots ,\hat{m}_{N}\right) ^{T}$, the noise 
\begin{equation}
\hat{\mathcal{N}}=\left( \sqrt{\kappa _{1}}\hat{N}_{1},\cdots ,\sqrt{\kappa
_{N}}\hat{N}_{N}\right) ^{T},
\end{equation}%
and the (microwave) torque 
\begin{equation}
\hat{\mathcal{T}}\equiv (\hat{T}_{1},\cdots ,\hat{T}_{N})^{T}=i\sum_{\lambda
}\int \hat{p}_{k,\mathrm{in}}^{\lambda }e^{-i\Omega _{k}^{\lambda }t}%
\mathcal{G}_{k}^{\lambda }\frac{dk}{\sqrt{2\pi }},  \label{torques}
\end{equation}%
with coupling $\mathcal{G}_{k}^{\lambda }=\left( g_{1}^{\lambda }(k),\dots
,g_{N}^{\lambda }(k)\right) ^{T}$. A local antenna such as metal-wire coils
close to each sphere \cite{MagnonDarkModes} can locally excite or detect its
dynamics, leading to the distributed torque $\hat{\mathcal{T}}\rightarrow 
\hat{\mathcal{T}}+\hat{\mathcal{T}}_{l},$ where $\hat{\mathcal{T}}_{l}=(\hat{%
P}_{1},\cdots ,\hat{P}_{N})^{T}$ and $\hat{P}_{i}$ is the local input
amplitude. The elements of the matrices $\tilde{\omega}$ and $\Sigma $ read 
\begin{align}
\tilde{\omega}_{jl}& =\delta _{jl}\left( \omega _{j}-i\frac{\kappa _{j}}{2}%
\right) , \\
\Sigma _{jl}& =\int_{0}^{\infty }\tilde{\Sigma}_{jl}(t)e^{i\omega _{l}t}dt.
\end{align}%
Inserting $\tilde{\Sigma}$, we obtain the self-energy 
\begin{equation}
\Sigma _{jl}=\sum_{\lambda }\int \frac{dk}{2\pi }\frac{g_{j}^{\lambda
}(k)g_{l}^{\lambda \ast }(k)}{\omega _{l}-\Omega _{k}^{\lambda }+i0^{+}}.
\label{Def:Sigma}
\end{equation}%
According to Eq. (\ref{EOM:Mags}), $\mathrm{\func{Re}}\Sigma $ modulates the
frequencies of each magnon by the other magnons (coherent coupling), while $%
\mathrm{\func{Im}}\Sigma $ changes the damping (dissipative coupling). We
discuss $\Sigma $ in more detail for a rectangular waveguide below.

\subsection{Collective modes}

\label{Sec:Quasiparticles} The coupling between magnets by photon exchange
in the waveguide gives rise to collective excitations. In the language of
quantum optics \cite%
{Molmer,subradiance1,subradiance2,subradiance3,subradiance4,subradiance5,Gardiner_chiral,input_output1}%
, Eq.~(\ref{EOM:Mags}) can be interpreted as a non-Hermitian Hamiltonian, $%
\hat{H}_{\mathrm{eff}}=\hbar \hat{\mathcal{M}}^{\dagger }\tilde{H}_{\mathrm{%
eff}}\hat{\mathcal{M}}$, with matrix 
\begin{equation}
\tilde{H}_{\mathrm{eff}}=\left( \tilde{\omega}+\Sigma \right) ,
\label{eff_Hamiltonian}
\end{equation}%
which (without input $\hat{\mathcal{T}})$ recovers the Heisenberg equation 
\cite%
{Molmer,subradiance1,subradiance2,subradiance3,subradiance4,subradiance5}.
Master equations lead to an effective non-Hermitian Hamiltonian by
exploiting the Monte Carlo wave-function method in quantum optics \cite%
{Molmer}. In general, any two systems coupled via continuous travelling
waves are dissipatively coupled.

The right and left eigenvectors of the non-Hermitian $\tilde{H}_{\mathrm{eff}%
}$ are not the same. Let the right eigenvectors of $\tilde{H}_{\mathrm{eff}}$
be $\{\psi _{\zeta }\}$ with corresponding eigenvalues $\{\nu _{\zeta }\}$
where $\zeta \in \{1,\dots ,N\}$ label the collective modes. It is also
convenient to define the right eigenvectors of $\tilde{H}_{\mathrm{eff}%
}^{\dagger }$ as $\{\phi _{\zeta }\}$ with corresponding eigenvalues $\{\nu
_{\zeta }^{\ast }\}$. Without degeneracies, i.e. $\forall _{\zeta \zeta
^{\prime }}$ $\nu _{\zeta }\neq \nu _{\zeta ^{\prime }}$, we have
bi-orthonormality $\psi _{\zeta }^{\dagger }\phi _{\zeta ^{\prime }}=\delta
_{\zeta \zeta ^{\prime }}$ after normalization. $\phi _{\zeta }^{\dagger }$
is a left eigenvector of $\tilde{H}_{\mathrm{eff}}$. The non-uniqueness of
the normalization condition does not affect the observables.

Defining matrices $\mathcal{L}=\left( \phi _{1},\dots ,\phi _{N}\right) $
and $\mathcal{R}=\left( \psi _{1},\dots ,\psi _{N}\right) $ in terms of left
and right eigenvectors, bi-orthonormality $\mathcal{R}^{\dagger }\mathcal{L}=%
\mathcal{L}^{\dagger }\mathcal{R}=I_{N}$, where $I_{N}$ is the $N\times N$
identity matrix, leads to 
\begin{equation}
\tilde{\omega}+\Sigma =\mathcal{R}\nu \mathcal{L}^{\dagger },
\label{effective2}
\end{equation}%
with matrix elements $\nu _{ij}=\left( \nu _{1},\dots ,\nu _{N}\right)
\delta _{ij}$. Defining 
\begin{equation}
\hat{\alpha}_{\zeta }=\phi _{\zeta }^{\dagger }\hat{\mathcal{M}},
\label{left_vector}
\end{equation}%
$\hat{\alpha}_{\zeta }$ annihilates a quasiparticle in a collective mode
with \textquotedblleft wave function\textquotedblright\ $\psi _{\zeta }$.
Substituting Eq.~(\ref{effective2}) into Eq.~(\ref{EOM:Mags}) leads to the
equation of motion 
\begin{equation}
\frac{d\hat{\alpha}_{\zeta }}{dt}=-i\nu _{\zeta }\hat{\alpha}_{\zeta }-\hat{%
\tau}_{\zeta }-\hat{N}_{\zeta },  \label{EOM:zeta}
\end{equation}%
where 
\begin{equation}
\hat{\tau}_{\zeta }=\phi _{\zeta }^{\dagger }\hat{\mathcal{T}};\ \ \hat{N}%
_{\zeta }=\phi _{\zeta }^{\dagger }\hat{\mathcal{N}}.  \label{torq:zeta}
\end{equation}%
The magnetization follows from the right eigenvectors 
\begin{equation}
\hat{\mathcal{M}}(t)=\sum_{\zeta }\hat{\alpha}_{\zeta }(t)\psi _{\zeta }.
\label{right_vector}
\end{equation}

\subsection{Photon scattering matrix}

\label{scattering_matrix}

The coupled set of magnets leads to collective excitations that affect the
transmission and reflection of input photons with frequency $\omega _{%
\mathrm{in}}$. The ensemble average $\left\langle {\cdots }\right\rangle $
of input mode $\lambda $ is
\begin{equation}
\left\langle \hat{p}_{k,\mathrm{in}}^{\lambda }\right\rangle =\sqrt{2\pi }%
A_{\lambda }\delta (k-k_{\lambda }),
\end{equation}%
where $A_{\lambda }$ is the amplitudes of the incoming microwave field and $%
k_{\lambda }$ is the positive wave vector satisfying $\Omega _{k_{\lambda
}}^{\lambda }=\omega _{\mathrm{in}}$. $\Omega _{k}^{\lambda }=\Omega
_{-k}^{\lambda }$ and we assume that $k_{\lambda }$ is unique, which is
satisfied in the absence of spatial modulations. The average of the torque
Eq.~(\ref{torq:zeta}) acting on mode $\zeta $ 
\begin{equation}
\left\langle \hat{\tau}_{\zeta }\right\rangle =i\sum_{\lambda }A_{\lambda
}e^{-i\omega _{\mathrm{in}}t}\mathcal{A}_{\zeta +}^{\lambda }.
\end{equation}%
The absorption coefficients 
\begin{equation}
\mathcal{A}_{\zeta \pm }^{\lambda }\equiv \phi _{\zeta }^{\dagger }\mathcal{G%
}_{\pm k_{\lambda }}^{\lambda }  \label{Absorption:Def}
\end{equation}%
are a linear combination of $g_{j}^{\lambda }$'s with weights given by the
left eigenvector. We argue below that the latter may be localized to only a
few magnets, such that a local coupling constant can dominate the global
absorption.

The average amplitude of mode $\zeta $ follows from Eq.~(\ref{EOM:zeta}). In
the steady state 
\begin{equation}
\left\langle \hat{\alpha}_{\zeta }(t)\right\rangle =\sum_{\lambda
}A_{\lambda }e^{-i\omega _{\mathrm{in}}t}\frac{\mathcal{A}_{\zeta
+}^{\lambda }}{\omega _{\mathrm{in}}-\nu _{\zeta }}.  \label{Amplitude:SS}
\end{equation}%
Mode $\zeta $ is resonantly excited when $\omega _{\mathrm{in}}=\mathrm{%
\mathrm{Re}}\nu _{\zeta }$ with spectral broadening $\mathrm{Im}\nu _{\zeta
} $. The photon output Eq.~(\ref{IO:Basic}) is 
\begin{equation}
\left\langle \hat{p}_{k,\mathrm{out}}^{\lambda }\right\rangle =\left\langle 
\hat{p}_{k,\mathrm{in}}^{\lambda }\right\rangle -i\sum_{\zeta }\mathcal{E}%
_{\zeta \pm }^{\lambda }\int \hat{\alpha}_{\zeta }(\tau )e^{i\Omega
_{k}^{\lambda }\tau }\frac{d\tau }{\sqrt{2\pi }},
\end{equation}%
with $+$ ($-$) sign for $k>0$ ($k<0$), while the emission coefficient 
\begin{equation}
\mathcal{E}_{\zeta \pm }^{\lambda }\equiv \mathcal{G}_{\pm k_{\lambda
}}^{\lambda \dagger }\psi _{\zeta }  \label{Emission:Def}
\end{equation}%
is a linear combination of couplings $g_{j}^{\lambda }$ weighted by the
right eigenvector. When the latter is localized, emission is governed by a
few magnetic moments and couplings between them. 

The coherent output 
\begin{equation}
\left\langle \hat{p}_{k,\mathrm{out}}^{\lambda }\right\rangle =\sqrt{2\pi }%
\sum_{\lambda ^{\prime }}\left[ S_{12}^{\lambda \lambda ^{\prime }}\delta
(k-k_{\lambda })+S_{22}^{\lambda \lambda ^{\prime }}\delta (k+k_{\lambda })%
\right] A_{\lambda ^{\prime }},
\end{equation}%
contains a transmission 
\begin{equation}
S_{12}^{\lambda \lambda ^{\prime }}(\omega _{\mathrm{in}})=\delta _{\lambda
\lambda ^{\prime }}-\frac{i}{v^{\lambda }(k_{\lambda })}\sum_{\zeta =1}^{N}%
\frac{\mathcal{E}_{\zeta +}^{\lambda }\mathcal{A}_{\zeta +}^{\lambda
^{\prime }}}{\omega _{\mathrm{in}}-\nu _{\zeta }}  \label{S12:Gen}
\end{equation}%
and a reflection amplitude 
\begin{equation}
S_{11}^{\lambda \lambda ^{\prime }}(\omega _{\mathrm{in}})=-\frac{i}{%
v^{\lambda }(k_{\lambda })}\sum_{\zeta =1}^{N}\frac{\mathcal{E}_{\zeta
-}^{\lambda }\mathcal{A}_{\zeta +}^{\lambda ^{\prime }}}{\omega _{\mathrm{in}%
}-\nu _{\zeta }},  \label{S11:Gen}
\end{equation}%
with photon group velocity 
\begin{equation}
v^{\lambda }(k)=\left\vert d\Omega _{k}^{\lambda }/dk\right\vert .
\label{Def:Vel}
\end{equation}%
$S_{21}$ and $S_{22}$ can be found respectively from $S_{11}$ and $S_{12}$
by the substitution $\mathcal{A}_{\zeta +}^{\lambda ^{\prime }}\rightarrow 
\mathcal{A}_{\zeta -}^{\lambda ^{\prime }}$. The (inter-band) scattering
amplitudes resonate at $N$ eigen frequencies of the collective magnetic
modes.

This result can be derived as well from scattering theory \cite%
{scattering_PRB,scattering_PRE,Fano,Mahan}.

\section{Rectangular waveguide}

\label{Sec:Rect}

We discuss here the coupling matrix $\Sigma $ for a rectangular waveguide
with cross-section from $(0,0)$ to $(a\geq b,b),$ with a detailed derivation
in Appendix~\ref{App:Sigma}. We use transverse mode indices $\lambda \equiv
\{n_{x},n_{y},\sigma \}$, in which integers $n_{x},n_{y}\geq 0$ are the
number of nodes of magnetic (or electric) field in the $\mathbf{x}$- and $%
\mathbf{y}$-directions, and $\sigma \in \{\mathrm{TE},\mathrm{TM}\}$ denotes
the polarization. The photon dispersion is \cite{Jackson} 
\begin{equation}
\Omega _{k}^{\lambda }=c\sqrt{k^{2}+\gamma _{\lambda }^{2}},
\end{equation}%
where $\gamma _{\lambda }\equiv \sqrt{(\gamma _{x}^{\lambda })^{2}+(\gamma
_{y}^{\lambda })^{2}}$ with $\gamma _{x}^{\lambda }=\pi n_{x}/a$ and $\gamma
_{y}^{\lambda }=\pi n_{y}/b$, does not depend on polarization index $\sigma $%
.

The diagonal elements of the coupling $\Sigma _{jj}$ in Eq.~(\ref{Def:Sigma}%
) represent self-interaction that shifts the frequencies by a small amount ($%
\mathrm{\func{Re}}\Sigma _{jj}\ll \omega _{j}$ as shown below) and describe
the radiative damping $\mathrm{\func{Im}}\Sigma _{jj}$, see Sec. \ref%
{one_sphere}. The non-diagonal elements $\Sigma _{i\neq l}$ couple different
magnets. With $\tilde{g}_{j}^{\lambda }(k)=-ig_{j}^{\lambda }(k)e^{-ikz_{j}} 
$, where $\mathrm{\func{Im}}\tilde{g}_{j}^{\lambda }(k)=0$ (see Appendix~\ref%
{App:Sigma}), we obtain an effective coupling 
\begin{equation}
\Sigma _{jl}=\sum_{\lambda }^{\mathrm{\func{Im}}k_{l}^{\lambda }=0}%
\begin{cases}
-i\frac{\Gamma _{L}+\Gamma _{R}}{2}-\delta \omega _{j}^{\lambda }, & j=l \\ 
-i\Gamma _{R}e^{ik_{l}^{\lambda }\left( z_{j}-z_{l}\right) }, & z_{j}>z_{l}
\\ 
-i\Gamma _{L}e^{ik_{l}^{\lambda }\left( z_{l}-z_{j}\right) }, & z_{j}<z_{l}%
\end{cases}%
,  \label{Sigma:Rect}
\end{equation}
that is modulated by a phase factor depending on the locations of the
magnets.

Here, the frequency shift for magnet $j$ by the photon band $\lambda $ reads 
\begin{equation}
\delta \omega _{j}^{\lambda }=\frac{\gamma \mu _{0}M_{s}V_{s}k_{c}}{ab}\sin
^{2}\left( \gamma _{x}^{\lambda }x_{j}\right) \cos ^{2}\left( \gamma
_{y}^{\lambda }y_{j}\right) ,
\end{equation}%
where $k_{c}$ is an upper cut-off for the wave numbers, which is typically
governed by high-frequency losses in the boundaries. For typical electron
relaxation time in copper, $\tau _{\mathrm{el}}=50$~fs ($\Omega _{c}\sim
2\pi \times 20$~THz) \cite{ElecRelax}, $k_{c}=2\pi /(\tau _{\mathrm{el}%
}c)\sim 10^{5}$~$\mathrm{m}^{-1}$ and $\delta \omega _{j}^{\lambda }\lesssim
2\pi \times 100$~MHz for $a\sim b\sim 2$~cm and the sphere radius of $0.5$%
~mm, which is much smaller than the Kittel mode frequency $\omega _{j}\sim
2\pi \times 10$~GHz. The inter-magnet coupling (suppressing various indices) 
\begin{equation}
\Gamma _{R}=\frac{\tilde{g}_{j}^{\lambda }\left( k_{l}^{\lambda }\right) 
\tilde{g}_{l}^{\lambda }\left( k_{l}^{\lambda }\right) }{v^{\lambda }\left(
k_{l}^{\lambda }\right) },
\end{equation}%
with group velocity Eq.~(\ref{Def:Vel}) 
\begin{equation}
v^{\lambda }(k)={c^{2}|k|}/{\Omega _{k}^{\lambda }},
\end{equation}%
and (positive) wave number of the photons emitted by the $l$-th magnet is 
\begin{equation}
k_{l}^{\lambda }=\sqrt{{\omega _{l}^{2}}/{c^{2}}-\gamma _{\lambda }^{2}}.
\end{equation}%
The summation in Eq. (\ref{Sigma:Rect}) is limited over the $\lambda $'s for
which $k_{l}^{\lambda }$ is real, i.e. the frequency of the $l$-th magnet is
larger than the $\lambda $-band edge. $\Gamma _{L}$ is obtained from $\Gamma
_{R}$ by $k_{l}^{\lambda }\rightarrow -k_{l}^{\lambda }$

For our rectangular waveguide, the couplings between magnets mediated by the
TM- and TE-photons are 
\begin{equation}
\tilde{g}_{j}^{\lambda }\left( \pm k_{l}^{\lambda }\right) |_{\mathrm{TM}%
}=G_{l}\frac{\gamma _{y}^{\lambda }}{\gamma _{\lambda }}\sin \left( \gamma
_{x}^{\lambda }x_{j}\right) \cos \left( \gamma _{y}^{\lambda }y_{j}\right) ,
\end{equation}%
and 
\begin{align}
\tilde{g}_{j}^{\lambda }\left( \pm k_{l}^{\lambda }\right) |_{\mathrm{TE}}&
=G_{l}\frac{ck_{l}^{\lambda }}{\omega _{j}}\frac{\gamma _{x}^{\lambda }}{%
\gamma _{\lambda }}\cos \left( \gamma _{y}^{\lambda }y_{j}\right)  \notag \\
& \times \left[ \sin \left( \gamma _{x}^{\lambda }x_{j}\right) \pm \frac{%
\gamma _{\lambda }^{2}}{k_{l}^{\lambda }\gamma _{x}^{\lambda }}\cos \left(
\gamma _{x}^{\lambda }x_{j}\right) \right] ,
\end{align}%
respectively, where 
\begin{equation}
G_{l}=\sqrt{\frac{\tilde{\gamma}\mu _{0}M_{s}V_{s}\omega _{l}}{ab}}.
\end{equation}%
For the TE modes, the magnon-photon coupling depends on the direction of
propagation. The chirality $\tilde{g}_{j}^{\lambda }\left( k_{l}^{\lambda
}\right) |_{\mathrm{TE}}\neq \tilde{g}_{j}^{\lambda }\left( -k_{l}^{\lambda
}\right) |_{\mathrm{TE}}$ is caused by a magnetic field that is not linearly
polarized, as indicated for $\{n_{x}=1,n_{y}=0\}$ mode in Fig.~\ref%
{fig:chiral}. When $z_{j}>z_{l}$ and the $j$-th magnet is located at a
position $x_{j}$ satisfying 
\begin{equation}
\cot \left( \frac{\pi x_{j}}{a}\right) =-\sqrt{\frac{a^{2}\omega _{l}^{2}}{%
\pi ^{2}c^{2}}-1},  \label{condition_chiral}
\end{equation}%
the magnon-photon coupling is fully chiral $\Sigma _{jl}=0$, so the $l$-th
magnet does not affect the $j$-th one. The coupling is also nonreciprocal, i.e.,
one magnet feels the dynamics of another, but not the other way around.

\begin{figure}[th]
{\includegraphics[width=8.5cm]{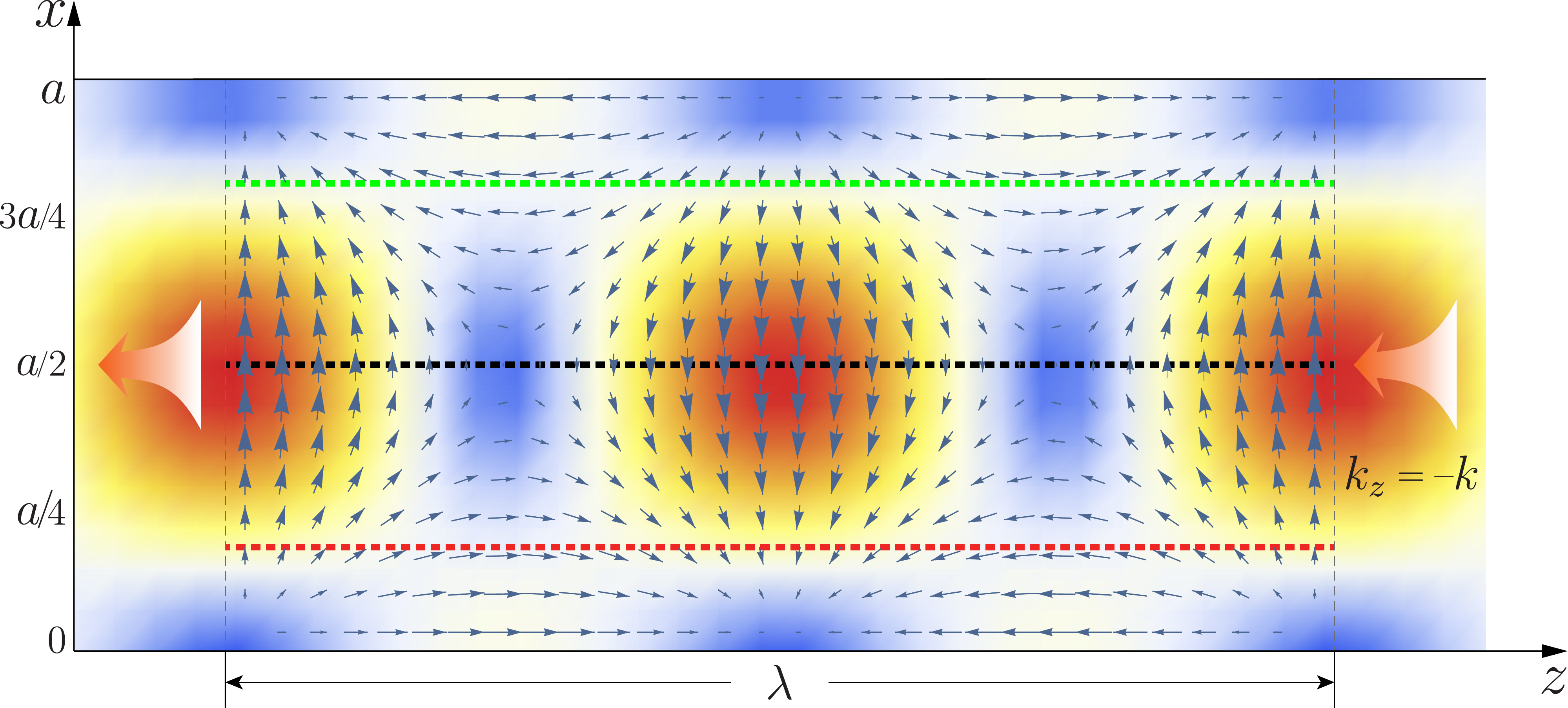}}
\caption{Snapshot of the spatial distribution of the AC magnetic field of
the lowest TE$_{10}$ mode in a rectangular wave guide propagating along the $%
-\mathbf{z}$-direction. The arrows indicate the direction and modulus of the
field. The latter is also indicated by the color shading, from zero (dark
blue) to maximum value (dark red). The vector field of modes along the $%
\mathbf{z}$-direction (not shown) is reversed. The green and red (black)
dotted lines indicate the locations at which the magnon-photon coupling is
chiral (non-chiral) for magnon frequency tuned to $\protect\omega _{l}=(2/%
\protect\sqrt{3})c\protect\pi /a$. On the red (green) line, the magnon mode
only couples to photons with positive (negative) linear momentum.}
\label{fig:chiral}
\end{figure}

When tuning the magnon frequency to below the bottom of all $\lambda $-bands
except for the lowest $\mathrm{TE}_{10}$ mode (the $\mathrm{TE}_{00}$ mode
does not exist), i.e. 
\begin{equation}
\frac{\pi }{a}<\frac{\omega _{l}}{c}<\left\{ \frac{\pi }{b},\frac{2\pi }{a}%
\right\},
\end{equation}%
we can freely tune the chirality. Fig.~\ref{fig:chiral} shows a snapshot of
the magnetic field for the lowest $\mathrm{TE}_{10}$ mode propagating along
the $-\mathbf{z}$-direction. For the modes along the $\mathbf{z}$-direction,
the local ellipticity is reversed. Solving Eq.~(\ref{condition_chiral}) with 
$\omega _{l}=(2/\sqrt{3})c\pi /a$, magnon-photon coupling is fully chiral
for magnets on the green and red dotted line. The chirality vanishes on the
center (black dotted) line and is partially chiral everywhere else. Spectral
overlap with TM-photons at higher frequencies would reduce the chirality.

\section{Microwave emission by magnetization dynamics}

\label{one_sphere}

Analogous to the spin pumping \cite%
{spin_pumping_electron,non_local,pumping_linear}, the transfer of energy and
angular momentum from magnons to photons implies radiative damping. In a
waveguide, this can be much larger than the intrinsic damping of a
high-quality magnet such as YIG \cite{RD2A,RD2,Magnon_radiation}. Radiative
damping also exists in free space, as derived in Appendix~\ref{App:RadFree},
but in the waveguide we can control its magnitude.

\subsection{Radiative damping}

In this section, we focus on a single magnet with (Kittel) frequency $\omega
_{m}$. The magnon lifetime broadening $\delta \omega =2\left( \alpha
_{G}+\alpha _{r}\right) \omega $, where $\alpha _{G}$ is the Gilbert damping
parameter and [see Eq. (\ref{Sigma:Rect})] \cite{Abrikosov,Fetter,Mahan,Haug}%
, 
\begin{equation}
\alpha _{r}=\frac{-\mathrm{\func{Im}}\Sigma }{\omega _{m}}=\sum_{\lambda }%
\frac{\left\vert g^{\lambda }\left( k^{\lambda }\right) \right\vert
^{2}+\left\vert g^{\lambda }\left( -k^{\lambda }\right) \right\vert ^{2}}{%
2c^{2}k^{\lambda }},  \label{Broadening}
\end{equation}%
where 
\begin{equation}
k^{\lambda }=\sqrt{\frac{\omega _{m}^{2}}{c^{2}}-\left( \frac{\pi n_{x}}{a}%
\right) ^{2}-\left( \frac{\pi n_{y}}{b}\right) ^{2}}.
\end{equation}

\begin{figure}[th]
\begin{center}
{\includegraphics[width=7.5cm]{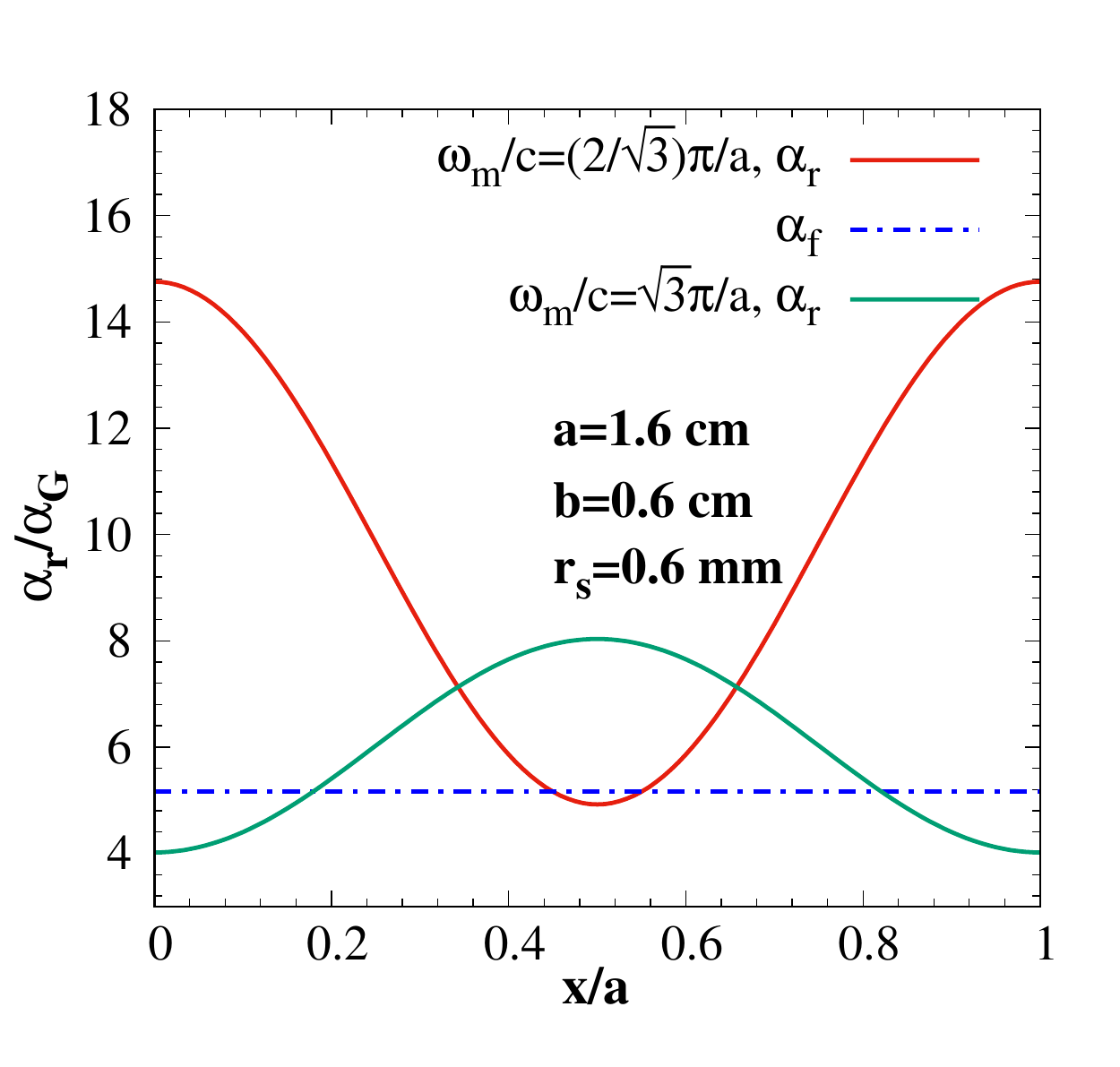}}
\end{center}
\caption{(Color online) Position-dependent radiative damping $\protect\alpha %
_{r}$ scaled by the intrinsic Gilbert damping $\protect\alpha _{G}$ of a YIG
sphere by the TE$_{10}$ mode of a rectangular waveguide\ for two magnetic
frequencies $\protect\omega _{m}$. The parameters including $a$ are specfied
in the text. The free space radiative damping $\protect\alpha _{f}$ from
Appendix~\protect\ref{App:RadFree} is also given (for the larger $\protect%
\omega _{m}=(2/\protect\sqrt{3})c\protect\pi /a$).}
\label{damping_x}
\end{figure}

We are mainly interested in the radiative damping of the lowest TE$_{10}$
mode of a rectangular waveguide. The mode amplitude and the associated
radiative damping do not depend on the $y$-coordinate. Results are plotted
in Fig.~\ref{damping_x} for $\omega _{m}=(2/\sqrt{3})c\pi /a$ where $a=1.6$%
~cm, $b=0.6$~cm, a magnetic sphere with radius $r_{s}=0.6$~mm and intrinsic
Gilbert damping $\alpha _{G}=5\times 10^{-5}$ \cite{Magnon_radiation} for
two frequencies. $\alpha _{r}$ depends strongly on $x$, but weaker when close
to the special position of chiral coupling, i.e., $x=a/3$ and $2a/3$ at $%
\omega _{m}=(2/\sqrt{3})c\pi /a$. The radiative dissipation in the waveguide
can be much larger than the viscous Gilbert damping as well as the radiative
damping in free space \cite{RD2A}, see Appendix~\ref{App:RadFree}, Eq.~(\ref%
{FreeRad:Res}): \textit{\ } \textit{\ } 
\begin{equation}
\alpha _{f}=\frac{\tilde{\gamma}\mu _{0}M_{s}V_{s}\omega _{m}^{2}}{6\pi c^{3}%
}.  \label{free}
\end{equation}%
$\alpha _{f}$ scales like $\omega _{m}^{2}$, and it can become larger than $%
\alpha _{r}$ at higher frequencies, because the photon density of states is
suppressed by the waveguide.

The broadening of the ferromagnetic resonance is not so sensitive to $g$'s chirality, but the
transmission is. In the $\lambda =\{1,0,\text{TE}\}$ mode the scattering
matrix in Eqs.~(\ref{S12:Gen}) and (\ref{S11:Gen}) reduces to 
\begin{equation}
S_{12}\left( \omega _{\mathrm{in}}\right) =\frac{\omega _{\mathrm{in}%
}-\omega _{m}+i\alpha _{G}\omega _{m}+i\left( \Gamma _{L}-\Gamma _{R}\right)
/2}{\omega _{\mathrm{in}}-\omega _{m}+i\alpha _{G}\omega _{m}+i\left( \Gamma
_{L}+\Gamma _{R}\right) /2},
\end{equation}%
where 
\begin{equation}
\Gamma _{R}\equiv \frac{\left\vert g^{\lambda }(k^{\lambda })\right\vert ^{2}%
}{v^{\lambda }(k^{\lambda })},\ \ \Gamma _{L}\equiv \frac{\left\vert
g^{\lambda }(-k^{\lambda })\right\vert ^{2}}{v^{\lambda }(k^{\lambda })}.
\end{equation}%
and $\alpha _{r}\omega _{m}=(\Gamma _{L}+\Gamma _{R})/2.$ When $\Gamma
_{L}=\Gamma _{R}$, the transmission amplitude drops at the resonance $\omega
_{\mathrm{in}}=\omega _{m}$ to a small value $\sim \alpha _{G}\omega
_{m}/\Gamma _{R}$. However, for full chirality with $\Gamma _{R}=0$, the
magnet does not absorb photons travelling towards the right and the
waveguide is transparent. When $\Gamma _{L}=0,$ on the other hand the
transmission probability is still unity, but the phase is shifted by $\pi $.

\subsection{Spatial chirality of dipolar field emission}

\label{linear_response} The AC magnetic field in the waveguide emitted by a
dynamical magnetic moment can be expressed by the linear response \cite%
{pumping_linear,non_local}, 
\begin{equation}
H_{\alpha }^{r}(\mathbf{r},t)=-\mu _{0}\int d\mathbf{r}^{\prime }dt^{\prime
}\chi _{\alpha \beta }(\mathbf{r}-\mathbf{r}^{\prime },t-t^{\prime
})M_{\beta }(\mathbf{r}^{\prime },t^{\prime }),
\end{equation}%
where the non-local inverse susceptibility $\chi _{\alpha \beta }$ is a
correlation function of the photon magnetic field $\mathbf{\hat{H}}$ 
\begin{equation}
\chi _{\alpha \beta }(\mathbf{r}-\mathbf{r}^{\prime },t-t^{\prime })=i\Theta
(t-t^{\prime })\left\langle \left[ \hat{H}_{\alpha }(\mathbf{r},t),\hat{H}%
_{\beta }(\mathbf{r}^{\prime },t^{\prime })\right] \right\rangle .
\end{equation}%
For the present system 
\begin{align}
& \chi _{\alpha \beta }(\pmb{\rho},z,\pmb{\rho}^{\prime },z^{\prime
};t-t^{\prime })  \notag \\
& =i\Theta (t-t^{\prime })\sum_{\lambda }\int \mathcal{H}_{k,\alpha
}^{\lambda }\left( \pmb{\rho}\right) \mathcal{H}_{k,\beta }^{\lambda \ast
}\left( \pmb{\rho}^{\prime }\right) e^{ik(z-z^{\prime })-i\Omega
_{k}^{\lambda }(t-t^{\prime })}\frac{dk}{2\pi }.
\end{align}%
Disregarding the small damping, $M_{\beta }(\mathbf{r}^{\prime },t^{\prime
})=M_{\beta }\left( \mathbf{r}^{\prime }\right) e^{i\omega _{m}(t-t^{\prime
})}$ and 
\begin{align}
H_{\alpha }^{r}(\mathbf{r},t)& =\mu _{0}\sum_{\lambda }\int d\mathbf{r}%
^{\prime }\int \frac{dk}{2\pi }\mathcal{H}_{k,\alpha }^{\lambda }(\pmb{\rho})%
\mathcal{H}_{k,\beta }^{\lambda \ast }(\pmb{\rho}^{\prime })  \notag \\
& \times e^{ik(z-z^{\prime })}\frac{1}{\omega _{m}-\Omega _{k}^{\lambda
}+i0^{+}}M_{\beta }(\mathbf{r}^{\prime },t).  \label{complex}
\end{align}%
By contour integration over $k$ for $z>z^{\prime }$ 
\begin{align}
H_{\alpha }^{r>}(\mathbf{r},t)& =-i\mu _{0}\sum_{\lambda }\frac{1}{%
v(k_{\lambda })}\mathcal{H}_{k_{\lambda },\alpha }^{\lambda }(\pmb{\rho}) 
\notag \\
& \times \int \mathcal{H}_{k_{\lambda },\beta }^{\lambda \ast }(\pmb{\rho}%
^{\prime })M_{\beta }(\mathbf{r}^{\prime },t)e^{ik_{\lambda }(z-z^{\prime
})}d\mathbf{r}^{\prime },  \label{eqn:out_of_phase1}
\end{align}%
and for $z<z^{\prime }$ 
\begin{align}
H_{\alpha }^{r<}(\mathbf{r},t)& =-i\mu _{0}\sum_{\lambda }\frac{1}{%
v(k_{\lambda })}\mathcal{H}_{-k_{\lambda },\alpha }^{\lambda }(\pmb{\rho}) 
\notag \\
& \times \int \mathcal{H}_{-k_{\lambda },\beta }^{\lambda \ast }(\pmb{\rho}%
^{\prime })M_{\beta }(\mathbf{r}^{\prime },t)e^{-ik_{\lambda }(z-z^{\prime
})}d\mathbf{r}^{\prime }.  \label{eqn:out_of_phase}
\end{align}%
When the coupling is chiral, the self-interaction magnetic field (for
equilibrium magnetization along $\mathbf{y)}$ becomes 
\begin{align}
\tilde{H}_{\alpha \in \{x,z\}}^{r}(\mathbf{r},t)& =\frac{\mu _{0}V_{s}}{%
2\omega _{m}}\sum_{\lambda }\frac{1}{v(k_{\lambda })}  \notag \\
& \times \left( \left\vert \mathcal{H}_{k_{\lambda },\alpha }(\pmb{\rho}%
)\right\vert ^{2}+\left\vert \mathcal{H}_{-k_{\lambda },\alpha }(\pmb{\rho}%
)\right\vert ^{2}\right) \frac{dM_{\alpha }(\mathbf{r},t)}{dt},
\end{align}%
which is out-of-phase with the local magnetization and therefore acts like
an additional and anisotropic Gilbert damping torque \cite{RD1,RD2A,RD2,RD3}.

The linear response formulation \cite{pumping_linear,non_local} helps to
understand the radiative damping: the precessing magnetization in a magnet
radiates dipolar magnetic field that is out-of-phase with the
magnetization. The self-interaction leads to a Gilbert damping-like torque.
This may be interpreted in terms of pumping of energy and angular momentum
into the microwave field. By substituting the linearized LLG equation \cite%
{Landau},
\begin{equation}
\frac{dM_{\alpha }}{dt}=\varepsilon _{\alpha \beta \delta }M_{\beta }\left( -%
\tilde{\gamma}\mu _{0}H_{\mathrm{eff},\delta }+\tilde{\gamma}\mu _{0}\tilde{H%
}_{\delta }^{r}+\frac{\alpha _{G}}{M_{s}}\frac{dM_{\delta }}{dt}\right) ,
\label{LLG1}
\end{equation}%
and the radiative damping is anisotropic 
\begin{equation}
\alpha _{\delta =\{x,z\}}^{\left( r\right) }=\frac{\mu _{0}^{2}V_{s}}{%
2\omega _{m}}\sum_{\lambda }\frac{\tilde{\gamma}M_{s}}{v(k_{\lambda })}%
\left( |\mathcal{H}_{k_{\lambda },\delta }(\pmb{\rho})|^{2}+|\mathcal{H}%
_{-k_{\lambda },\delta }(\pmb{\rho})|^{2}\right) .
\end{equation}%
Linearizing Eq.~(\ref{LLG1}) and substituting $M_{\alpha }\propto
e^{-i\omega t}$ yields 
\begin{eqnarray}
i\omega M_{x}+(\tilde{\gamma}\mu _{0}H_{\mathrm{eff},y}-i\omega \alpha
_{z}^{\left( r\right) }-i\omega \alpha _{G})M_{z} &=&0,  \notag \\
i\omega M_{z}-(\tilde{\gamma}\mu _{0}H_{\mathrm{eff},y}-i\omega \alpha
_{x}^{\left( r\right) }-i\omega \alpha _{G})M_{x} &=&0,
\end{eqnarray}%
and the quadratic equation 
\begin{equation}
\omega ^{2}+i\omega \tilde{\gamma}\mu _{0}H_{\mathrm{eff},y}(\alpha
_{x}^{\left( r\right) }+\alpha _{z}^{\left( r\right) }+2\alpha _{G})-(\tilde{%
\gamma}\mu _{0}H_{\mathrm{eff},y})^{2}=0.
\end{equation}%
Therefore 
\begin{equation*}
\alpha _{G}^{\mathrm{eff}}\left( \omega \right) \approx \alpha _{G}+(\alpha
_{x}^{\left( r\right) }+\alpha _{z}^{\left( r\right) })/2=\alpha _{G}+\alpha
_{r}\left( \omega \right) ,
\end{equation*}%
consistent with the equation of motion approach. The full damping tensor can
in principle be reconstructed by computing the dependence of $\alpha _{r}$
on the magnetization direction.

\section{Magnon hydrogen molecule}

\label{two_sphere}

The interaction is non-local since the photons emitted by one magnet are
reabsorbed by another magnet, which is a basically classical phenomenon (see
Appendix \ref{Coup:Class}), even though we derived it by the Heisenberg
equation of motion in Sec. \ref{Sec:EOMs} and discussed in more detail for a
rectangular waveguide in Sec. \ref{Sec:Rect}. The classical electrodynamics
in Appendix \ref{Coup:Class} becomes tedious for multiple magnets, so we focus
in the following on the quantum description of two magnets, turning to the
magnet chain in Sec.~\ref{many_sphere}.

\subsection{Collective mode}

We consider the transmission of a single waveguide mode with input amplitude 
$A_{\lambda }$ and frequency $\omega _{\mathrm{in}}$. In the following we
suppress the mode index $\lambda $, i.e., $A_{\lambda }=A$, $S_{12}\equiv
S_{12}^{\lambda \lambda }$, $v\equiv v^{\lambda }(k_{\lambda })$. $k\equiv
k_{\lambda }$ is the wave vector of the incoming photons and $\mathcal{G}%
=(g_{1},g_{2})^{T}\equiv \mathcal{G}_{k_{\lambda }}^{\lambda }$ is the
vector of couplings $g_{j}$ of the $j$-th magnet.

The two spheres are oriented along the waveguide with $\pmb{\rho}_{1}=%
\pmb{\rho}_{2}$, and $d=z_{2}>z_{1}=0$. The magnetic input field amplitude
at the spheres differs by the phase $kd$. According to Sec. \ref{Sec:Rect} 
\begin{equation}
\mathcal{G}=-ig_{0}%
\begin{pmatrix}
1 & e^{ikd}%
\end{pmatrix}%
^{T},
\end{equation}%
where $g_{0}$ is real. The frequency shift and radiative damping of the
resonances in both magnets are the same and we absorb them into the complex
frequencies $\omega _{1},\omega _{2}$. The Hamiltonian matrix then reads 
\begin{align}
\tilde{H}_{\mathrm{eff}}& =\tilde{\omega}+\Sigma  \notag \\
& =%
\begin{pmatrix}
\omega _{1}-i\alpha _{G}\omega _{1}-i\frac{\Gamma _{L}+\Gamma _{R}}{2} & 
-i\Gamma _{L}e^{ikd} \\ 
-i\Gamma _{R}e^{ikd} & \omega _{2}-i\alpha _{G}\omega _{2}-i\frac{\Gamma
_{L}+\Gamma _{R}}{2}%
\end{pmatrix}%
.  \label{matrix_two_spheres}
\end{align}%
We assume $\omega _{1}\approx \omega _{2}\approx \omega _{\mathrm{in}}$, but
allow them to vary in a window small enough that $\Gamma (\omega
_{1})\approx \Gamma (\omega _{2})\approx \Gamma (\omega _{\mathrm{in}})$.

As discussed in Sec. \ref{Sec:Quasiparticles}, the eigenvectors of $\tilde{%
\omega}+\Sigma $, namely $\{\psi _{+},\psi _{-}\}$, with corresponding
eigenvalues $\{\nu _{+},\nu _{-}\}$, and eigenvectors of $\left( \tilde{%
\omega}+\Sigma \right) ^{\dagger }$, namely $\{\phi _{+},\phi _{-}\}$
contain relevant information of the observables. Here 
\begin{eqnarray}
\nu _{+}+\nu _{-} &=&\left( \omega _{2}+\omega _{1}\right) \left( 1-i\alpha
_{G}\right) -i\left( \Gamma _{L}+\Gamma _{R}\right) ,  \notag \\
\nu _{+}-\nu _{-} &=&\sqrt{\left( \omega _{2}-\omega _{1}\right) ^{2}\left(
1-i\alpha _{G}\right) ^{2}-4\Gamma _{L}\Gamma _{R}e^{2ikd}},
\label{MagAmpTwo}
\end{eqnarray}%
correspond to two resonant frequencies and linewidths. Assuming $1-i\alpha
_{G}\approx 1$, 
\begin{align}
\psi _{\pm }& \approx X_{\pm }%
\begin{pmatrix}
\Delta \pm \sqrt{\Delta ^{2}-4\Gamma _{L}\Gamma _{R}e^{2ikd}} \\ 
2i\Gamma _{R}e^{ikd}%
\end{pmatrix}%
,  \notag \\
\phi _{\pm }& \approx Y_{\pm }%
\begin{pmatrix}
2i\Gamma _{R}e^{-ikd} \\ 
\Delta \mp \sqrt{\Delta ^{2}-4\Gamma _{L}\Gamma _{R}e^{-2ikd}}%
\end{pmatrix}%
,
\end{align}
with the detuning $\omega _{2}-\omega _{1}=\Delta $. The normalization
factors 
\begin{equation}
X_{\pm }Y_{\pm }^{\ast }=\frac{\pm i}{4\Gamma _{R}e^{ikd}\sqrt{\Delta
^{2}-4\Gamma _{L}\Gamma _{R}e^{2ikd}}}
\end{equation}%
are chosen such that $\phi _{\pm }^{\dagger }\psi _{\pm }=1$.

The absorption coefficient [Eq. (\ref{Absorption:Def})] 
\begin{align}
\mathcal{A}_{\pm }& =\phi _{\pm }^{\dagger }\mathcal{G}  \notag \\
& =-ig_{0}Y_{\pm }^{\ast }e^{ikd}\left[ \Delta -2i\Gamma _{R}\mp \sqrt{%
\Delta ^{2}-4\Gamma _{L}\Gamma _{R}e^{2ikd}}\right] ,
\end{align}%
and the excited magnetization can be written as 
\begin{equation}
\left\langle \hat{\mathcal{M}}\right\rangle =\left\langle \hat{\alpha}%
_{+}(t)\right\rangle \psi _{+}+\left\langle \hat{\alpha}_{-}(t)\right\rangle
\psi _{-},
\end{equation}%
with amplitudes [Eq.~(\ref{Amplitude:SS})] 
\begin{equation}
\left\langle \hat{\alpha}_{\pm }(t)\right\rangle =Ae^{-i\omega _{\mathrm{in}%
}t}\frac{\mathcal{A}_{\pm }}{\omega _{\mathrm{in}}-\nu _{\pm }}.
\label{excitation_2}
\end{equation}

\subsection{Directional pumping of magnons}

\label{two_chiral} For zero detuning the resonant input $\omega _{\mathrm{in}%
}=\omega _{1}=\omega _{2}=\omega _{m}$ drives the magnetization of each
sphere into a coherent state $\left\langle \hat{m}\right\rangle $ with some
thermal noise, see Sec.~\ref{scattering_matrix}. The ratio of the coherent
amplitudes 
\begin{equation}
\Lambda \overset{\triangle }{=}\left\vert \frac{\left\langle \hat{m}%
_{1}\right\rangle }{\left\langle \hat{m}_{2}\right\rangle }\right\vert =%
\sqrt{\frac{\Gamma _{L}}{\Gamma _{R}}}\left\vert \frac{\left\langle \alpha
_{+}\right\rangle X_{+}-\left\langle \alpha _{-}\right\rangle X_{-}}{%
\left\langle \alpha _{+}\right\rangle X_{+}+\left\langle \alpha
_{-}\right\rangle X_{-}}\right\vert ,
\end{equation}%
does not depend on time. With Eq.~(\ref{excitation_2}) 
\begin{equation}
\Lambda =\left\vert \frac{2\alpha _{G}\omega _{m}+\Gamma _{R}+\Gamma
_{L}\left( 1-e^{2ikd}\right) }{2\alpha _{G}\omega _{m}+\Gamma _{L}-\Gamma
_{R}}\right\vert .
\end{equation}%
$\Lambda ^{2}$ is the ratio of the coherent magnon numbers (also refer to
the results of the master equation below).

An imbalanced excitation $\Lambda \gg 1$ can exist even without chirality,
i.e., when $\Gamma _{L}=\Gamma _{R}=\Gamma $: $\Lambda =\left\vert 1+\Gamma
\left( 2-e^{2ikd}\right) /\left( 2\alpha _{G}\omega _{m}\right) \right\vert
, $ which is caused by the directional excitation of a waveguide, but
depends strongly on the parameters. When $\Gamma _{R}\rightarrow 0$ and $%
\Gamma _{L}\gg \alpha _{G}\omega _{m}$ we obtain the universal $\Lambda
\approx 5-4\cos (2kd)$ . When $kd=n\pi /2$ with $n$ being odd integer, $%
\Lambda =9$, and a ratio of the excited magnon numbers of $\Lambda
^{2}\approx 81$. When $\Gamma _{L}=\Gamma _{R}-2\alpha _{G}\omega _{m},$ $%
\left\vert \left\langle \hat{m}_{2}\right\rangle \right\vert =0$ and $%
\Lambda $ diverges: magnet $2$ cannot be excited because the input and
emitted photons from the other magnet interfere destructively. This limit
can be realized by shifting the magnets in the waveguide and/or tuning the
applied field.

Magnons can also be excited locally by small local antennas with negligible
cross talk \cite{magnonics1,magnonics2,magnonics3,magnonics4,Yu1,Yu2}. An
imbalanced magnon excitation can be detected by the same antenna, as
pioneered in the cavity experiment \cite{MagnonDarkModes}. We can model
local drives by adding source terms to the equation of motion Eq. (\ref%
{EOM:Mags}), 
\begin{equation}
\frac{d\left\langle \hat{\mathcal{M}}\right\rangle }{dt}=-i\left( \tilde{%
\omega}+\Sigma \right) \left\langle \hat{\mathcal{M}}\right\rangle +%
\begin{pmatrix}
\left\langle \hat{P}_{1}(t)\right\rangle \\ 
\left\langle \hat{P}_{2}(t)\right\rangle%
\end{pmatrix},
\end{equation}%
where $\hat{P}_{i}$ are the local magnetic field amplitudes and we ignored
the dissipation caused by the local antennas, for simplicity. When $%
\left\langle \hat{P}_{i}(t)\right\rangle =iPe^{-i\omega _{\mathrm{in}}t}$,
where $P$ is real, 
\begin{equation}
\Lambda =\left\vert \frac{\Gamma _{L}+\Gamma _{R}+2\alpha _{G}\omega
_{m}-2\Gamma _{L}e^{ikd}}{\Gamma _{L}+\Gamma _{R}+2\alpha _{G}\omega
_{m}-2\Gamma _{R}e^{ikd}}\right\vert .  \label{independent}
\end{equation}%
In contrast to the waveguide drive discussed above, the excitation is
balanced when $\Gamma _{L}=\Gamma _{R}$. $\Lambda \neq 1$ then requires
chiral coupling, e.g., when $\Gamma _{R}=0$, $\Lambda ^{2}\approx 5-4\cos
(kd)\leq 9$. This imbalance is caused by the pumping of the first magnet by
the second magnet without back-action.

The coherent and dissipative components of the coupling emerge in the
equation of motion from the commutator of the Hamiltonian with the magnon
operator. Their different physical meanings can be understood best by the
master equation \cite{Molmer,input_output1,Gardiner_chiral,chiral_emitter}.
To this end we divide the non-Hermitian Hamiltonian into the Hermitian $\hat{%
H}_{h}$ and non-Hermitian $\hat{H}_{nh}$ parts as 
\begin{equation}
\hat{H}_{\mathrm{eff}}=(\hat{H}_{\mathrm{eff}}+\hat{H}_{\mathrm{eff}%
}^{\dagger })/2+(\hat{H}_{\mathrm{eff}}-\hat{H}_{\mathrm{eff}}^{\dagger })/2,
\label{eqn:division}
\end{equation}%
with the first and second terms representing the Hermitian and non-Hermitian
parts, respectively. For the magnon hydrogen molecule 
\begin{align}
\hat{H}_{h}& =\sum_{i=1,2}\omega _{i}\hat{m}_{i}^{\dagger }\hat{m}_{i}+i%
\frac{\Sigma _{12}+\Sigma _{21}^{\ast }}{2}\hat{m}_{1}^{\dagger }\hat{m}_{2}
\notag \\
& +i\frac{\Sigma _{21}+\Sigma _{12}^{\ast }}{2}\hat{m}_{1}\hat{m}%
_{2}^{\dagger }, \\
\hat{H}_{nh}& =-i\sum_{i=1,2}\frac{\delta \omega _{m}}{2}\hat{m}%
_{i}^{\dagger }\hat{m}_{i}+\frac{\Sigma _{12}-\Sigma _{21}^{\ast }}{2}\hat{m}%
_{1}^{\dagger }\hat{m}_{2}  \notag \\
& +\frac{\Sigma _{21}-\Sigma _{12}^{\ast }}{2}\hat{m}_{1}^{\dagger }\hat{m}%
_{2},
\end{align}%
with $\delta \omega _{m}=\Gamma _{R}+\Gamma _{L}+2\alpha _{G}\omega _{m}$, $%
\Sigma _{12}=-i\Gamma _{L}e^{ikd}$ and $\Sigma _{21}=-i\Gamma _{R}e^{ikd}$.
The coherent and dissipative contribution cause different collective
dampings \cite{Molmer,input_output1,Gardiner_chiral,chiral_emitter}. The
master equation for the density operator of magnon $\hat{\rho}$ \cite%
{Molmer,input_output1,Gardiner_chiral,chiral_emitter}, 
\begin{align}
\partial _{t}\hat{\rho}& =i\left[ \hat{\rho},\hat{H}_{h}\right] +\sum_{i}%
\frac{\delta \omega _{m}}{2}\hat{\mathcal{L}}_{ii}\hat{\rho}+i\frac{\Sigma
_{12}-\Sigma _{21}^{\ast }}{2}\hat{\mathcal{L}}_{12}\hat{\rho}  \notag \\
& +i\frac{\Sigma _{21}-\Sigma _{12}^{\ast }}{2}\hat{\mathcal{L}}_{21}\hat{%
\rho},
\end{align}%
in which $\mathcal{L}_{ij}\hat{\rho}=2\hat{m}_{j}\hat{\rho}\hat{m}%
_{i}^{\dagger }-\hat{m}_{i}^{\dagger }\hat{m}_{j}\hat{\rho}-\hat{\rho}\hat{m}%
_{i}^{\dagger }\hat{m}_{j}$ is a relaxation operator (Lindblad
super-operator), while $\delta \omega _{m}$ and $i(\Sigma _{12}-\Sigma
_{21}^{\ast })/2$ are the self and collective decay rates, respectively. For
perfect chiral coupling $\Sigma _{21}=0$ and at resonance, the master
equation in the rotating frame and $\hat{m}(t)=\tilde{m}e^{-i\omega _{%
\mathrm{in}}t}$ gives for the slowly varying envelopes $\tilde{m}_{1,2}$ 
\begin{equation}
\frac{\partial }{\partial t}%
\begin{pmatrix}
\langle \tilde{m}_{1}\rangle \\ 
\langle \tilde{m}_{2}\rangle%
\end{pmatrix}%
=%
\begin{pmatrix}
-\delta \omega _{m}/2 & -i\Sigma _{12} \\ 
0 & -\delta \omega _{m}/2%
\end{pmatrix}%
\begin{pmatrix}
\langle \tilde{m}_{1}\rangle \\ 
\langle \tilde{m}_{2}\rangle%
\end{pmatrix}%
+%
\begin{pmatrix}
-iP \\ 
-iP%
\end{pmatrix}%
,  \label{master1}
\end{equation}%
where the average $\langle \hat{O}(t)\rangle =\langle \hat{O}\hat{\rho}%
(t)\rangle $, and 
\begin{align}
&\frac{\partial }{\partial t}%
\begin{pmatrix}
\langle \tilde{m}_{1}^{\dagger }\tilde{m}_{1}\rangle \\ 
\langle \tilde{m}_{2}^{\dagger }\tilde{m}_{2}\rangle \\ 
\langle \tilde{m}_{1}^{\dagger }\tilde{m}_{2}\rangle \\ 
\langle \tilde{m}_{1}\tilde{m}_{2}^{\dagger }\rangle%
\end{pmatrix}%
 =%
\begin{pmatrix}
iP & -iP & 0 & 0 \\ 
0 & 0 & iP & -iP \\ 
0 & -iP & iP & 0 \\ 
-iP & 0 & 0 & iP%
\end{pmatrix}%
\begin{pmatrix}
\langle \tilde{m}_{1}\rangle \\ 
\langle \tilde{m}_{2}\rangle \\ 
\langle \tilde{m}_{1}^{\dagger }\rangle \\ 
\langle \tilde{m}_{2}^{\dagger }\rangle%
\end{pmatrix}
\notag \\
& +%
\begin{pmatrix}
-\delta \omega _{m} & 0 & -i\Sigma _{12} & i\Sigma _{12}^{\ast } \\ 
0 & -\delta \omega _{m} & 0 & 0 \\ 
0 & \Sigma _{12}^{\ast } & -\delta \omega _{m} & 0 \\ 
0 & 0 & 0 & -\delta \omega _{m}%
\end{pmatrix}%
\begin{pmatrix}
\langle \tilde{m}_{1}^{\dagger }\tilde{m}_{1}\rangle \\ 
\langle \tilde{m}_{2}^{\dagger }\tilde{m}_{2}\rangle \\ 
\langle \tilde{m}_{1}^{\dagger }\tilde{m}_{2}\rangle \\ 
\langle \tilde{m}_{1}\tilde{m}_{2}^{\dagger }\rangle%
\end{pmatrix}%
.  \label{master2}
\end{align}%
The coherent amplitude and associated magnon number\ (accumulation) obey
different equations. $P$ drives the coherent amplitude via Eq.~(\ref{master1}%
), while the dissipative coupling in Eq.~(\ref{master2}) causes collective
damping of the magnon numbers. We thus show that the master equation
approach is equivalent to the input-output theory: Eqs.~(\ref{master1})
and (\ref{master2}) recover the previous results for $\Lambda $ and $\Lambda
^{2}$ in Eq.~(\ref{independent}).

\section{Magnon chain}

\label{many_sphere}

The imbalance of the magnon distribution is enhanced when more magnets are
added to the waveguide. Let us consider a chain of $N$ identical magnets
located on a line $\pmb{\rho}_{\forall i}=\pmb{\rho}$ at equal distance $%
z_{j+1}-z_{j}=d\ \left( 0<j<N\right) $ as realized already for $N=7$ (but in
a closed cavity) \cite{MagnonDarkModes}. We study the eigenvectors and
eigenvalues of the non-Hermitian matrix 
\begin{widetext}
\begin{equation}
\tilde{H}_{\mathrm{eff}}=%
\begin{pmatrix}
\omega _{m}-i\alpha _{G}\omega _{m}-i\frac{\Gamma _{R}+\Gamma _{L}}{2} & 
-i\Gamma _{L}e^{ikd} & -i\Gamma _{L}e^{2ikd} & \dots & -i\Gamma
_{L}e^{(N-1)ikd} \\ 
-i\Gamma _{R}e^{ikd} & \omega _{m}-i\alpha _{G}\omega _{m}-i\frac{\Gamma
_{R}+\Gamma _{L}}{2} & -i\Gamma _{L}e^{ikd} & \dots & -i\Gamma
_{L}e^{(N-2)ikd} \\ 
-i\Gamma _{R}e^{2ikd} & -i\Gamma _{R}e^{ikd} & \omega _{m}-i\alpha
_{G}\omega _{m}-i\frac{\Gamma _{R}+\Gamma _{L}}{2} & \dots & -i\Gamma
_{L}e^{(N-3)kd} \\ 
\vdots & \vdots & \vdots & \ddots & \vdots \\ 
-i\Gamma _{R}e^{i(N-1)kd} & -i\Gamma _{R}e^{i(N-2)kd} & -i\Gamma
_{R}e^{i(N-3)kd} & \dots & \omega _{m}-i\alpha _{G}\omega _{m}-i\frac{\Gamma
_{R}+\Gamma _{L}}{2}%
\end{pmatrix}%
,  \label{H_matrix}
\end{equation}%
\end{widetext}
where we dropped the $\mathrm{TE}_{10}$ mode index $\lambda $ and 
\begin{equation}
k=\sqrt{\frac{\omega _{m}^{2}}{c^{2}}-\left( \frac{\pi }{a}\right) ^{2}}.
\end{equation}%
The photons emitted by magnet $j$ to the right are in our perturbative and
adiabatic approach seen equivalently and instantaneously by all magnets on
the right but with a phase factor $e^{ik|z_{j}-z_{l}|},$ and analogously for
the magnets to the left.

The photon-mediated interaction generates a band structure with generalized
Bloch states labelled $\zeta \in \{1,\dots ,N\}$ with right eigenvectors $%
\{\psi _{\zeta }\}$ and corresponding eigenvalues $\{\nu _{\zeta }\}$, 
\begin{equation}
(\nu _{\zeta }-\tilde{H}_{\mathrm{eff}})\psi _{\zeta }=0.
\end{equation}%
The real part of $\nu _{\zeta }$ is the resonance frequency of the $\zeta $%
-mode and the imaginary part its lifetime. The eigenvectors of $\tilde{H}_{%
\mathrm{eff}}^{\dagger }$, $\phi _{\zeta }$ with eigenvalue $\nu _{\zeta
}^{\ast }$ are related to $\psi _{\zeta }$ by a parity-time reversal
operation when the spectrum is not degenerate, which is the case for the
simple chain considered here. Let $\mathcal{T}$ be the complex conjugation
and 
\begin{equation}
\mathcal{P}=%
\begin{pmatrix}
0 & 0 & \dots & 0 & 1 \\ 
0 & 0 & \dots & 1 & 0 \\ 
\vdots & \vdots & \ddots & \vdots & \vdots \\ 
0 & 1 & \dots & 0 & 0 \\ 
1 & 0 & \dots & 0 & 0%
\end{pmatrix}%
\end{equation}%
exchanges the magnets $1\leftrightarrow N$, $2\leftrightarrow N-1$ and so
on, akin to the inversion operation. However, $\mathcal{P}$ does not act on
the waveguide and is therefore not a parity operator of the whole system.
Clearly, $\mathcal{P}^{2}=\mathcal{T}^{2}=1$. $\mathcal{P}$ interchanges $%
\Gamma _{L\leftrightarrow R}\ $in Eq. (\ref{H_matrix}), which is equivalent
to the transpose operation, i.e. $\tilde{H}_{\mathrm{eff}}^{T}=\mathcal{P}%
\tilde{H}_{\mathrm{eff}}\mathcal{P},$ while $\tilde{H}_{\mathrm{eff}%
}^{\dagger }=\mathcal{P}\mathcal{T}\tilde{H}_{\mathrm{eff}}\mathcal{T}%
\mathcal{P}$ and 
\begin{equation}
\tilde{H}_{\mathrm{eff}}^{\dagger }\mathcal{P}\mathcal{T}\psi _{\zeta }=\nu
_{\zeta }^{\ast }\mathcal{P}\mathcal{T}\psi _{\zeta },
\end{equation}%
implying that 
\begin{equation}
\phi _{\zeta }=\mathcal{P}\mathcal{T}\psi _{\zeta }.
\label{left_right_relation}
\end{equation}%
We chose a normalization 
\begin{equation}
\psi _{\zeta }^{T}\mathcal{P}\psi _{\zeta }=1  \label{normalization}
\end{equation}%
such that $\phi _{\zeta }^{\dagger }\psi _{\zeta }=1$. Thus, we can describe the
dynamics in terms of only the right eigenvectors $\psi _{\zeta }$.

The magnets interact with the photons (again suppressing indices) by the
phase vector 
\begin{equation}
\mathcal{G}=-i\sqrt{\Gamma _{R}v}\left( 1,e^{ikd},\dots ,e^{i(N-1)kd}\right)
^{T}.
\end{equation}%
The emission amplitude $\mathcal{E}_{\zeta }=\mathcal{G}^{\dagger }\psi
_{\zeta }=i\sqrt{\Gamma _{R}v}\tilde{\psi}_{\zeta }(k),$ where we defined
the discrete Fourier transform 
\begin{equation}
\tilde{\psi}_{\zeta }(k)=\left( 1,e^{-ikd},\dots ,e^{-i(N-1)kd}\right)
^{T}\psi _{\zeta }.
\end{equation}%
The absorption amplitude $\mathcal{A}_{\zeta }=\phi _{\zeta }^{\dagger }%
\mathcal{G}$ is related to the emission by 
\begin{equation}
\mathcal{A}_{\zeta }=e^{i(N-1)kd}\mathcal{E}_{\zeta }.
\end{equation}%
The global transmission [cf. Eq. (\ref{S12:Gen})] 
\begin{equation}
S_{12}(\omega _{\mathrm{in}})=1-i\Gamma _{R}e^{i(N-1)kd}\sum_{\zeta }\frac{%
\tilde{\psi}_{\zeta }^{2}(k)}{\omega _{\mathrm{in}}-\nu _{\zeta }},
\end{equation}%
is governed by the right eigenvectors. The total coherent magnetization of
the array 
\begin{equation}
\left\langle \hat{\mathcal{M}}(t)\right\rangle =A\sqrt{\Gamma _{R}v}%
e^{-i\omega _{\mathrm{in}}t}e^{i(N-1)kd}\sum_{\zeta }\frac{\tilde{\psi}%
_{\zeta }(k)}{\omega _{\mathrm{in}}-\nu _{\zeta }}\psi _{\zeta }
\end{equation}%
is proportional to the amplitude of the incoming photons $A$ (introduced in
Sec. \ref{scattering_matrix}).

Magnons can be flexibly excited and detected by local antennas that interact
only with one magnet \cite{MagnonDarkModes}. With local input at frequency $%
\omega _{\mathrm{in}}$, $\left\langle \hat{\mathcal{T}}_{l}(t)\right\rangle
=ie^{-i\omega _{\mathrm{in}}t}\left( P_{1},P_{2},\cdots ,P_{N}\right) ^{T}$, 
\begin{equation}
\left\langle \hat{\mathcal{M}}(t)\right\rangle =-i\sum_{\zeta }\frac{(%
\mathcal{P}\psi _{\zeta })^{T}\left\langle \hat{\mathcal{T}}%
_{l}(t)\right\rangle }{\omega _{\mathrm{in}}-\tilde{\omega}_{m}-\gamma
_{\zeta }}\psi _{\zeta }.
\end{equation}%
Note that $(\mathcal{P}\psi _{\zeta })^{T}=(\psi _{\zeta ,N},\psi _{\zeta
,N-1},\cdots ,\psi _{1})$. When an edge state ${\zeta _{\ast }}$ exists, say
on the right with large $\psi _{\zeta ,N}$, the antenna array with
controlled phase difference $\phi $, i.e. $\left\langle \hat{\mathcal{T}}%
_{l}(t)\right\rangle =\exp [{-i\mathrm{\mathrm{Re}}}\left( {\gamma _{\zeta
_{\ast }}}\right) {t}]iP(1,e^{i\phi },\cdots ,e^{i(N-1)\phi })^{T}$, can
excite a large magnetization at the right edge, where it can be detected by
the same local antenna as pointed out in the accompanying Letter \cite%
{PRL_submission}.

We see that the excitation of magnetization is determined by the
eigenvectors $\psi _{\zeta }$ and their eigenvalues $\nu _{\zeta }$, which
are studied numerically and analytically below, with special attention for
superradiant and subradiant modes, i.e. those with the largest and smallest
radiation rates, respectively.

\subsection{Numerical results}

\label{Sec:Numerical} We present and analyze numerical results for the
collective modes of the dissipatively-coupled magnon chain. As before, $%
a=1.6 $~cm, $b=0.6$~cm, $r_{s}=0.6$~mm, and $\alpha _{G}=5\times 10^{-5}$ 
\cite{Magnon_radiation}. Typically, $\omega _{m}/c=\sqrt{3}\pi /a$
corresponding to the photon momentum $k=\sqrt{2}\pi /a$, so only the lowest
TE$_{10}$ mode contributes. The magnetic chain is parallel to the waveguide
and shifted from the chiral line to modulate the chirality $\Gamma
_{R}/\Gamma _{L}=1,0.5,0.25,$ where $\Gamma _{L}/(2\pi )\in (0,20)$~MHz. We
choose $N=80$ magnetic spheres and $kd=\pi /5$. So $d=a/(5\sqrt{2})\approx
0.2$~cm, and the total length of the magnon chain is $Nd\approx 18$~cm. This
is much longer than our choice in the accompanying Letter \cite%
{PRL_submission}. While such a long chain is experimentally impractical, the
results are not qualitatively different and emphasize our message.

\begin{figure}[th]
\begin{center}
{\includegraphics[width=8.3cm]{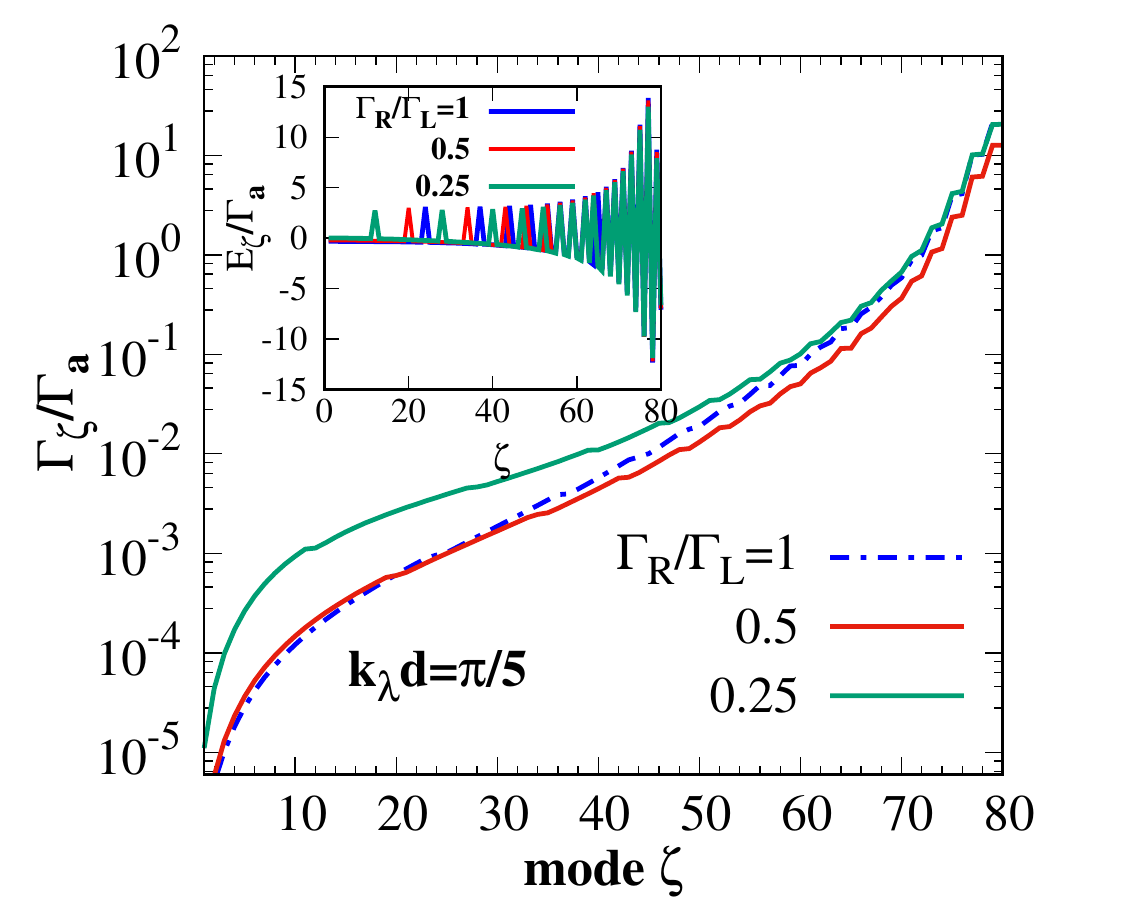}}
\end{center}
\caption{(Color online) Imaginary ($\Gamma _{\protect\zeta }$) and real
(inset, $E_{\protect\zeta }$) parts of the eigenvalues $(\protect\nu _{%
\protect\zeta }-\protect\omega _{m})$ of the non-Hermitian Hamiltonian [Eq.~(%
\protect\ref{H_matrix})], scaled by the individual damping rate $\Gamma _{a}$%
. $k_{\protect\lambda }d=\protect\pi /5$ and $N=80$. $\Gamma _{R}/\Gamma
_{L}=1$, 0.5 and 0.25, respectively. $E_{\protect\zeta }$ oscillates as a
function of $\protect\zeta $ and $\Gamma _{\protect\zeta }$ in a
non-systematic manner.}
\label{fig:brightdark_2}
\end{figure}

Fig.~\ref{fig:brightdark_2} is a plot of the imaginary ($\Gamma _{\zeta }$)
and real ($E_{\zeta }$) parts of $\nu _{\zeta }-\omega _{m}$ as a function
of mode number $\zeta $, scaled by the local dissipation rate $\Gamma
_{a}=\alpha _{G}\omega _{m}+(\Gamma _{L}+\Gamma _{R})/2$. The mode nnumbers $%
\zeta =\{1,2,...,N\}$ are ordered by magnitudes of $\Gamma _{\zeta }$. 
When $\Gamma _{R}=\Gamma _{L}$ (non-chiral case) and $\zeta \approx 80$ ($%
\zeta \lesssim 10$), the decay rates are larger (smaller) than the local $%
\Gamma _{a}$, indicating superradiance (subradiance). The decay rates of the
most-superradiant states $\sim \Gamma _{a}N/4$ can simply be enhanced by
increasing the number of magnets. The decay of the most-subradiant states $%
\sim \Gamma _{a}\zeta ^{2}/N^{3}$ \cite%
{subradiance1,subradiance2,subradiance3,subradiance4,subradiance5} are found
at the lower band edge. The value of the magnon energy shifts $E_{\zeta }$
in the inset of Fig.~\ref{fig:brightdark_2} are enhanced to a peak around
the boundary between sub- and superradiance ($\Gamma _{\zeta }\approx \Gamma
_{a}$). $E_{\zeta }$ and $\Gamma _{\zeta }$ have not simple functional
relationship, which is reflected by the oscillations (peaks) that look
erratic for small mode numbers. The energy shift of the most-subradiant
states is very small, but it can be as large as $\sim 10\Gamma _{a}$ for the
superradiant ones, roughly proportional to the number of magnets. The
largest energy shift $2\pi \times 100$~MHz is still small compared to $%
\omega _{m},$ which justifies the on-shell approximation for $\Gamma _{L}$
and $\Gamma _{R}$. $E_{\zeta }$ oscillates with $\zeta $ between positive
and negative values. A\textbf{\ }chiral coupling with $\Gamma _{R}/\Gamma
_{L}=0.5$ and 0.25 does not strongly change the above features, such the
decay rates of the most-subradiant states $\sim \Gamma _{a}\zeta ^{2}/N^{3}$%
\textit{.}

\begin{figure}[th]
\begin{center}
{\includegraphics[width=8.4cm]{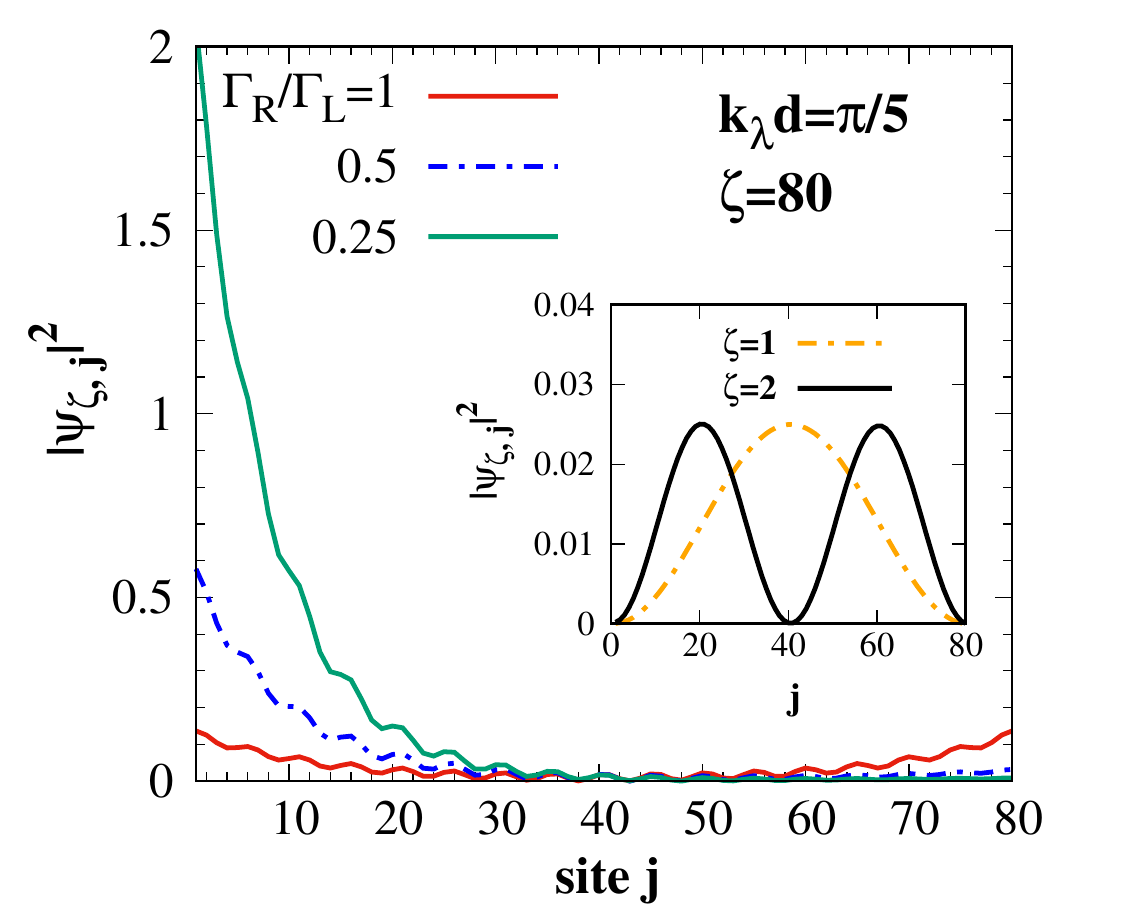}}
\end{center}
\caption{(Color online) Intensity distributions of magnons $|\protect\psi _{%
\protect\zeta ,j}|^{2}$ in magnetic spheres labeled by $j$ for the most
superadiant and subradiant (inset) states for chiralities $\Gamma
_{R}/\Gamma _{L}=1$, 0.5, and 0.25, respectively.}
\label{wavefunction}
\end{figure}

The intensity distributions $|\psi _{\zeta ,j}|^{2}$ of modes $\zeta =1,2,80$
over the chain $j=\{1,2,\cdots ,N\}$ are shown in Fig.~\ref{wavefunction}.
When $\Gamma _{R}=\Gamma _{L}$ for the non-chiral case, the
most-superradiant state is enhanced at both edges of the magnon chain (the
red solid curve). The most-subradiant states are standing waves $\sim
\left\vert \sin (\zeta \pi j/N)\right\vert $ delocalized over the whole
chain, but have small amplitudes at the edges (see the inset of Fig.~\ref%
{wavefunction}).


Partially chiral coupling does not affect the amplitude distributions of the
most-subradiant states. The symmetric distribution of the most-superradiant
states relative to the center of the chain $\Gamma _{R}=\Gamma _{L}$ becomes
increasingly skewed, i.e., the dynamics is enhanced at one edge only.
Particularly, when $\Gamma _{R}<\Gamma _{L}$ ($\Gamma _{R}>\Gamma _{L}$),
the edge state is localized at the left (right) side. When the radiation to
the left is stronger than to the right, the magnets on the left side
experience more radiation. On the other hand, the magnets in the middle of
the chain are part of a standing wave with destructive interference in the
average. A larger chirality $\Gamma _{R}/\Gamma _{L}$ consequently mainly
affects the edge states.

\subsection{Analytical analysis}

The rich features of the collective motion in the most-sub- and
super-radiant states can be accessed analytically in some special limits 
\cite{subradiance2}. To this end we search for linear combinations of the
magnon operators $\hat{\alpha}$ that satisfy $d\hat{\alpha}/dt=-i\nu \hat{%
\alpha}+\text{(noise)}$ as discussed in Sec.~\ref{Sec:Quasiparticles}.

We can rewrite the equation of motion for the magnetization, Eq.~(\ref%
{EOM:Mags}), as
\begin{equation}
\frac{d}{dt}(\hat{m}_{1},\hat{m}_{2},\cdots ,\hat{m}_{\delta })^{T}=-i\tilde{%
H}_{\mathrm{eff}}(\hat{m}_{1},\hat{m}_{2},\cdots ,\hat{m}_{\delta })^{T}.
\end{equation}%
\textit{\ }Inserting Eq.~(\ref{H_matrix}) for the magnon chain leads to 
\begin{align}
-\left( \frac{d\hat{m}_{\delta }}{dt}\right) _{c}& =\frac{\Gamma _{R}+\Gamma
_{L}}{2}\hat{m}_{\delta }+\Gamma _{R}\sum_{j<\delta }e^{ikd\left( \delta
-j\right) }\hat{m}_{j}  \notag \\
& +\Gamma _{L}\sum_{j>\delta }e^{ikd\left( j-\delta \right) }\hat{m}_{j}.
\label{EOM_0}
\end{align}%
where we dropped the noise term and the self-interaction $\propto i\omega 
\hat{m}_{\delta }$ that only contributes a constant, but does not affect the
eigenmodes. Inserting a trial Bloch state with complex momentum $\kappa $, 
\begin{equation}
\hat{m}_{\delta }\rightarrow \hat{\Psi}_{\kappa }=\frac{1}{\sqrt{N}}%
\sum_{j=1}^{N}e^{i{\kappa }z_{j}}\hat{m}_{j}  \label{extended_Bloch}
\end{equation}%
into Eq.~(\ref{EOM_0}) leads to 
\begin{equation}
\left( \frac{d\hat{\Psi}_{\kappa }}{dt}\right) _{c}=-i\omega _{\kappa }\hat{%
\Psi}_{\kappa }-\Gamma _{L}g_{\kappa }\hat{\Psi}_{k}+\Gamma _{R}h_{\kappa }%
\hat{\Psi}_{-k},  \label{EOM_Bloch_states}
\end{equation}%
with complex dispersion relation 
\begin{equation}
\omega _{\kappa }=-i\frac{\Gamma _{R}}{2}\frac{1+e^{i({\kappa }+k)d}}{1-e^{i(%
{\kappa }+k)d}}+i\frac{\Gamma _{L}}{2}\frac{1+e^{i({\kappa }-k)d}}{1-e^{i({%
\kappa }-k)d}},  \label{dispersion}
\end{equation}%
and `leakage' parameters 
\begin{equation}
g_{\kappa }=\frac{1}{1-e^{i({\kappa }-k)d}},\ \ h_{\kappa }=\frac{e^{i({%
\kappa }+k)Nd}}{1-e^{i({\kappa }+k)d}}.
\end{equation}%
Eq.~(\ref{EOM_Bloch_states}) is a closed equation for the unknown $\kappa $.
Only when the terms $g_{\kappa },h_{\kappa }$ in Eq.~(\ref{EOM_Bloch_states}%
) vanish, $\hat{\Psi}_{\kappa }$ is a proper solution. The leakage and
reflection at the edges mixes $\hat{\Psi}_{\kappa }$ with the plane waves $%
\hat{\Psi}_{k}\ $and $\hat{\Psi}_{-k}$, which renders the problem
non-trivial.

In general, the field operator $\hat{\alpha}$ should be a superposition of
frequency-degenerate Bloch waves. For the simple chain, two states with $%
\kappa $ and $\kappa ^{\prime }$ should suffice, provided 
\begin{equation}
\omega _{\kappa }=\omega _{\kappa ^{\prime }},  \label{first}
\end{equation}%
which leads to $({\Gamma _{R}-\Gamma _{L}+2\omega _{\kappa }})/(\Gamma
_{R}+\Gamma _{L}+2\omega _{\kappa })=-e^{i(\kappa ^{\prime }+\kappa )d}$.
Trying $\hat{\alpha}=g_{\kappa ^{\prime }}\hat{\Psi}_{\kappa }-g_{\kappa }%
\hat{\Psi}_{\kappa ^{\prime }}$, gives 
\begin{equation}
\left( \frac{d\hat{\alpha}}{dt}\right) _{c}=-i\omega _{\kappa }\hat{\alpha}%
+\Gamma _{R}\left( g_{\kappa ^{\prime }}h_{\kappa }-g_{\kappa }h_{\kappa
^{\prime }}\right) \hat{\Psi}_{-k},
\end{equation}%
which is the desired equation when 
\begin{equation}
g_{\kappa }h_{\kappa ^{\prime }}=g_{\kappa ^{\prime }}h_{\kappa }.
\label{second}
\end{equation}%
Eq.~(\ref{second}) is an $N$-th-order polynomial equation in $e^{i\kappa
_{\zeta }d}$ with $N$ roots. Since we have $N$ magnets and modes in the
non-interacting limits, its solutions cover all eigenvalues of the
interacting system. Eqs.~(\ref{first}) and (\ref{second}) suffice to
determine the complex unknown variables $\kappa $ and $\kappa ^{\prime }$. The wave
function and energies of collective mode can then be expanded as $\hat{\alpha%
}_{\zeta }=\sum_{j}\phi _{\zeta ,j}^{\ast }\hat{m}_{j}$ and with 
\begin{equation}
\hat{\alpha}=\frac{1}{\sqrt{N}}\sum_{j=1}^{N}(g_{\kappa ^{\prime
}}e^{ikz_{j}}-g_{\kappa }e^{ik^{\prime }z_{j}})\hat{m}_{j},
\end{equation}%
we obtain 
\begin{equation}
\phi _{j}^{\ast }=g_{\kappa ^{\prime }}e^{ikz_{j}}-g_{\kappa }e^{ik^{\prime
}z_{j}}.
\end{equation}%
Using the relation between the left and right eigenvectors [Eq.~(\ref%
{left_right_relation})], $\phi ^{\ast }=P\psi $, 
\begin{align}
\psi _{j}& \propto g_{\kappa ^{\prime }}e^{i\kappa z_{N-j}}-g_{\kappa
}e^{i\kappa ^{\prime }z_{N-j}}, \\
\nu & =\tilde{\omega}_{m}+\omega _{\kappa },  \label{solutions}
\end{align}%
with $z_{j}=(j-1)d$ and the normalization of $\psi _{\zeta }$ is given by
Eq.~(\ref{normalization}). For $\Gamma _{L}=\Gamma _{R}$ we find $\kappa
^{\prime }=-\kappa $ \cite{subradiance2}.

The imaginary part of $\omega _{\kappa }=\omega _{\kappa ^{\prime }}$
corresponds to the radiative damping of the mode $\zeta $. The super-radiant
modes with $\mathrm{\func{Im}}\omega _{\kappa }\gg \Gamma _{R},\Gamma _{L}$
are near $\kappa \approx \pm k$, i.e. complex momenta $\kappa =k_{0}+\eta $
and $\kappa ^{\prime }=-k_{0}+\eta ^{\prime }$ with small complex numbers $%
\eta $ and $\eta ^{\prime }$, which have to be calculated numerically. The
imaginary part of $\eta $ and $\eta ^{\prime }$ are reciprocal skin depths
of the edge states addressed in Sec.~\ref{Sec:Numerical}.

Near the minima of $\omega _{\kappa }$, around say ${\kappa }={\kappa }%
_{\ast }$, we expect sub-radiant modes. Minimizing Eq. (\ref{dispersion})
leads to 
\begin{align}
{\kappa }_{\ast }d& =\arcsin \frac{\Gamma _{R}-\Gamma _{L}}{\sqrt{\Gamma
_{R}^{2}+\Gamma _{L}^{2}-2\Gamma _{R}\Gamma _{L}\cos (2kd)}}  \notag \\
& -\arctan \frac{\Gamma _{R}-\Gamma _{L}}{(\Gamma _{R}+\Gamma _{L})\tan (kd)}%
.
\end{align}%
The $\arcsin $ is a two-valued function and hence we search for two extremal
points in the first Brillouin zone $[-\pi /d,\pi /d]$. $\kappa _{\ast }$ and
the corresponding $\kappa _{\ast }^{\prime }$ do not yet satisfy the
eigenvalue equation Eq.~(\ref{second}). Trying $\kappa =\kappa _{\ast
}+\delta $ and $\kappa ^{\prime }=\kappa _{\ast }-\delta $ leads to 
\begin{equation}
e^{2i\delta Nd}=\frac{\cos ({\kappa }_{\ast }d)-\cos [(k+\delta )d]}{\cos ({%
\kappa }_{\ast }d)-\cos [(k-\delta )d]}.  \label{delta:eq}
\end{equation}%
For $\left\vert \delta d\right\vert \ll 1$ 
\begin{equation}
\delta \approx \frac{\xi \pi }{Nd}\left[ 1-\frac{i}{N}\frac{\sin (kd)}{\cos (%
{\kappa }_{\ast }d)-\cos (kd)}\right] ,
\end{equation}%
where $\xi =\{1,2,\cdots \}$, leading to eigenfunctions 
\begin{align}
\psi _{\xi ,j}& \approx -2i\frac{e^{i\kappa _{\ast }z_{N-j}}}{1-e^{i(\kappa
_{\ast }-k)d}}\sin (\delta _{\xi }z_{N-j}),  \notag \\
\omega _{\xi }& =\omega _{{\kappa }_{\ast }}+\frac{\sin (kd)}{\cos ({\kappa }%
_{\ast }d)-\cos (kd)}\frac{\Gamma _{R}(\delta _{\xi }d)^{2}/2}{1-\cos [(k+{%
\kappa }_{\ast })d]},
\end{align}%
that are symmetric even for chiral coupling, because sub-radiant modes do
not efficiently couple to the waveguide. These results also explain the
standing-wave feature and scaling law of the radiative lifetime of these
states.

\section{Discussion and Conclusion}

\label{summary} In conclusion, we find and report the consequences of chiral
and dissipative coupling of small magnets to guided microwaves. We predict a
rich variety of physical phenomena, such as directional photon emission and
magnon imbalanced pumping and super-(sub-)radiance of collective magnon modes.
Polarization-momentum locking of the electromagnetic field inside a
rectangular waveguide and conservation of angular momentum are the physical
mechanisms behind chiral magnon-photons interaction. Chirality can be tuned
via the positions of the magnetic spheres inside the waveguide and applied
static magnetic fields. We develop the theory starting with a single magnet
and demonstrate strong radiative damping. Loading the waveguide with two or
more magnets causes nonreciprocal tunable coupling between different
magnetic spheres. We predict chirality-dependent large magnon amplitudes at
the edges of long chains with superradiance. We also reveal subradiant
eigenstates, which are standing waves with small amplitude at the edges,
that depend only weakly on chirality and therefore scale as different
systems without chirality \cite%
{subradiance1,subradiance2,subradiance3,subradiance4,subradiance5,subradiance6}%
.

The magnetic chain in a waveguide is also a new platform to study
non-Hermitian physics \cite%
{non_hermitian0,non_hermitian1,non_hermitian2,non_hermitian3,non_hermitian4}%
. The rich magnon-photon dynamics suggests several lines of future research.
Tunable waveguides allow manipulation of the local density of photon states
and linewidth for each collective mode~\cite{Magnon_radiation}, while
arrangements of the magnetic spheres into rings, lattices or random geometry
promise a new \textquotedblleft magnon chemistry\textquotedblright . Some
non-Hermitian Hamiltonians may result in topological phases, a hot topic in
condensed matter physics~\cite%
{non_hermitian5,non_hermitian6,non_hermitian7,non_hermitian8,non_hermitian9}%
. The non-Bloch-wave behavior of eigenstates of a chiral magnon-photon
system can cause a non-Hermitian skin effect and a non-Bloch bulk-boundary
correspondence. The non-linear dynamics of a chirally vs. non-chirally
coupled magnon-photon system can be accessed by the photon statistics of the
waveguide to specify the entanglement of sub- and super-radiant states~\cite%
{subradiance5}.

\begin{acknowledgments}
	This work is financially supported by the Nederlandse Organisatie voor Wetenschappelijk Onderzoek (NWO) as well as JSPS KAKENHI Grant No. 26103006. We would like to thank Yu-Xiang Zhang and Bi-Mu Yao for helpful discussions.
\end{acknowledgments}

\begin{appendix}

\section{Dissipative coupling}

\label{App:Sigma} Here we derive the radiative damping and dissipative
coupling between identical magnets in a rectangular waveguide by photons in
both TM and TE modes by explicitly calculating Eq.~(\ref{Def:Sigma}). For
simplicity, we drop the explicit dependence on $\lambda $ and $k$, i.e. $%
\Omega \equiv \Omega _{k}^{\lambda }$ and $g_{j}\equiv g_{j}^{\lambda }(k)$.

The magnetic field of the TM modes \cite{Jackson} 
\begin{align}
\mathcal{H}_{x}& =\sqrt{\frac{2\hbar \Omega }{\mu _{0}ab}}\frac{\gamma _{y}}{%
\gamma }\sin \left( \gamma _{x}x\right) \cos \left( \gamma _{y}y\right) , 
\notag \\
\mathcal{H}_{y}& =-\sqrt{\frac{2\hbar \Omega }{\mu _{0}ab}}\frac{\gamma _{x}%
}{\gamma }\cos \left( \gamma _{x}x\right) \sin \left( \gamma _{y}y\right) ,
\end{align}%
with both $n_{x},n_{y}>0$, and of the TE modes 
\begin{align}
\mathcal{H}_{z}& =-i\sqrt{\frac{\eta \hbar \Omega }{\mu _{0}ab}}\frac{%
c\gamma }{\Omega }\frac{|k|}{k}\cos \left( \gamma _{x}x\right) \cos \left(
\gamma _{y}y\right) ,  \notag \\
\mathcal{H}_{x}& =\sqrt{\frac{\eta \hbar \Omega }{\mu _{0}ab}}\frac{c|k|}{%
\Omega }\frac{\gamma _{x}}{\gamma }\sin \left( \gamma _{x}x\right) \cos
\left( \gamma _{y}y\right) ,  \notag \\
\mathcal{H}_{y}& =\sqrt{\frac{\eta \hbar \Omega }{\mu _{0}ab}}\frac{c|k|}{%
\Omega }\frac{\gamma _{y}}{\gamma }\cos \left( \gamma _{x}x\right) \sin
\left( \gamma _{y}y\right) ,
\end{align}%
in which $\eta =2-\delta _{n_{x},0}-\delta _{n_{y},0}$ and at least one $%
n_{x},n_{y}>0$. According to Eq.~(\ref{coupling}) 
\begin{equation}
g_{j}^{\mathrm{TM}}=i\sqrt{\frac{\tilde{\gamma}\mu _{0}M_{s}V_{s}\Omega }{ab}%
}\frac{\gamma _{y}}{\gamma }e^{ikz_{j}}\sin \left( \gamma _{x}x_{j}\right)
\cos \left( \gamma _{y}y_{j}\right) ,  \label{CoupTM}
\end{equation}%
and%
\begin{align}
g_{j}^{\mathrm{TE}}& =i\sqrt{\frac{\eta \tilde{\gamma}\mu
_{0}M_{s}V_{s}\Omega }{ab}}\frac{c|k|}{\Omega }\frac{\gamma _{x}}{\gamma }%
e^{ikz_{j}}\cos \left( \gamma _{y}y_{j}\right)  \notag \\
& \times \left[ \sin \left( \gamma _{x}x_{j}\right) +\frac{\gamma ^{2}}{%
k\gamma _{x}}\cos \left( \gamma _{x}x_{j}\right) \right] .  \label{CoupTE}
\end{align}%
At large $|k|$, these couplings increase proportionally to $\sqrt{|k|}$
because the magnetic field scales with the square-root of the photon energy.
The magnon-magnon coupling in Eq. (\ref{Def:Sigma}) then becomes%
\begin{align}
\Sigma _{jl}& =\frac{\tilde{\gamma}\mu _{0}M_{s}V_{s}}{ab}%
\sum_{n_{x},n_{y}}\sin \left( \gamma _{x}x_{j}\right) \cos \left( \gamma
_{y}y_{j}\right)  \notag \\
& \times \sin \left( \gamma _{x}x_{l}\right) \cos \left( \gamma
_{y}y_{l}\right) \left( \frac{\gamma _{x}^{2}}{\gamma ^{2}}\mathcal{I}_{%
\mathrm{TM}}+\frac{\gamma _{y}^{2}}{\gamma ^{2}}\mathcal{I}_{\mathrm{TE}%
}\right) ,
\end{align}%
where the $\mathcal{I}_{\sigma }$ summarize the TM and TE contributions. Here%
\begin{equation}
\mathcal{I}_{\mathrm{TM}}=\int \frac{\Omega e^{ik(z_{j}-z_{l})}}{\omega
_{l}-\Omega +i0^{+}}\frac{dk}{2\pi }.
\end{equation}%
The ultraviolet divergence for $z_{j}=z_{l}$ can be removed by introducing a
cut-off momentum $k_{c}$ that parametrizes dissipation in the metal
boundaries at high frequencies. 
\begin{equation}
\mathcal{I}_{\mathrm{TM}}|_{j=l}=\int_{-\infty }^{\infty }\frac{dk}{2\pi }%
\frac{\omega _{l}}{\omega _{l}-\Omega +i0^{+}}-2k_{c}\delta _{jl}.
\label{ITM:1}
\end{equation}
Using the Cauchy's relation ${1}/(x+i0^{+})=\mathcal{P}(1/{x})-i\pi \delta (x)$,
where $\mathcal{P}$ is the principle value, 
\begin{equation}
\mathcal{I}_{\mathrm{TM}}|_{j=l}=-i{\omega _{l}^{2}}/(c^{2}k_{l})-2k_{c},
\end{equation}%
where $k_{l}=\sqrt{{\omega _{l}^{2}}/{c^{2}}-\gamma ^{2}}$ with $\omega
_{l}\geq \gamma c$. The divergence of the imaginary part at the band edge $%
\omega _{l}\approx c\gamma $ is a harmless van Hove singularity.

When $z_{j}>z_{l}$ and $\gamma \left\vert z_{j}-z_{l}\right\vert >1$, the
photon mode with negative wave number is evanescent and cannot affect
another magnet that is not in immediate proximity. The integral then
simplifies to 
\begin{equation*}
\mathcal{I}_{\mathrm{TM}}|_{z_{j}>z_{l}}=\int_{-\infty }^{\infty }\frac{dk}{%
2\pi }\frac{\Omega e^{ik(z_{j}-z_{l})}}{\omega _{l}-\Omega +i0^{+}}\approx 
\frac{i\omega _{l}^{2}e^{ik_{l}(z_{j}-z_{l})}}{c^{2}k_{l}},
\end{equation*}%
consistent with Ref.~\cite{Jackson}, Sec~\ref{linear_response}, and our
numerical calculations. The restriction $2\pi (z_{j}-z_{l})\gtrsim {\min }%
\{a,b\}$ (or $k_{l}(z_{j}-z_{l})>1$) requires that for our system $%
z_{j}-z_{l}\gtrsim 0.2$~cm for ${\min }\{a,b\}\sim 1$~cm, which we assume to
be the case in the following. For $z_{j}<z_{l}$, a similar result holds with 
$z_{j}-z_{l}\rightarrow z_{l}-z_{j}$.

For TE modes 
\begin{equation}
\mathcal{I}_{\mathrm{TE}}=\int \frac{dk}{2\pi }\frac{c^{2}e^{ik(z_{j}-z_{l})}%
}{\Omega \left( \omega _{l}-\Omega +i0^{+}\right) }\left[ k+\frac{\gamma ^{2}%
}{\gamma _{x}}\cot (\gamma _{x}x_{j})\right] ^{2}.
\end{equation}%
We obtain 
\begin{equation}
\mathcal{I}_{\mathrm{TE}}|_{j=l}=-i\frac{1}{k_{l}}\left[ k_{l}^{2}+\frac{%
\gamma ^{4}}{\gamma _{x}^{2}}\cot ^{2}(\gamma _{x}x_{j})\right] -2k_{c}.
\end{equation}%
When $\gamma \left\vert z_{j}-z_{l}\right\vert >1$, 
\begin{equation}
\mathcal{I}_{\mathrm{TE}}|_{z_{j}-z_{l}}=-i\frac{e^{ik_{l}|z_{j}-z_{l}|}}{%
k_{l}}\left[ k_{l}+\frac{\gamma ^{2}}{\gamma _{x}}\cot (\gamma _{x}x_{j})%
\right] ^{2}.
\end{equation}

\section{Free space radiation damping}

\label{App:RadFree}

Here we drive the radiation damping of the Kittel mode of a single magnet in
free space addressed in Sec. \ref{general}. The magnetic field can be
expanded 
\begin{equation}
\mathbf{H}(\mathbf{r})=\sum_{\sigma }\int \frac{d^{3}k}{(\sqrt{2\pi })^{3}}%
\sqrt{\frac{\hbar \Omega _{k}}{2\mu _{0}}}\mathbf{e}_{\mathbf{k}}^{\sigma
}\left( e^{i\mathbf{k}.\mathbf{r}}\hat{p}_{\mathbf{k}}+e^{-i\mathbf{k}.%
\mathbf{r}}\hat{p}_{\mathbf{k}}^{\dagger }\right) .
\end{equation}%
The frequency $\Omega _{k}=ck$ and the two polarization vectors are 
\begin{equation}
\mathbf{e}_{\mathbf{k}}^{1}=\frac{\left( k_{z},0,-k_{x}\right) }{\sqrt{%
k_{x}^{2}+k_{z}^{2}}},\ \ \mathbf{e}_{\mathbf{k}}^{2}=\frac{k_{y}\mathbf{k}%
-k^{2}\mathbf{y}}{k\sqrt{k_{x}^{2}+k_{z}^{2}}}.
\end{equation}%
The coupling with a magnet with equilibrium magnetization along $\mathbf{y}$ 
\begin{equation}
g_{\mathbf{k}}^{\sigma }=\sqrt{\frac{\mu _{0}\Omega _{k}}{2}\frac{\gamma
M_{s}V_{s}}{2}}\left( ie_{\mathbf{k},x}^{\sigma }-ie_{\mathbf{k},z}^{\sigma
}\right) .
\end{equation}%
The broadening of the ferromagnetic resonance is given by Fermi's Golden
Rule [analogous to Eq. (\ref{Broadening})], 
\begin{equation}
\Delta \omega =\sum_{\sigma }\int \frac{d^{3}k}{(2\pi )^{2}}\delta \left(
\omega _{m}-\Omega _{k}\right) \left\vert g_{\mathbf{k}}^{\sigma
}\right\vert ^{2},
\end{equation}%
where $\omega _{m}$ is the magnon frequency. $\left\vert g_{\mathbf{k}%
}^{\sigma }\right\vert ^{2}$ can be simplified by the relations 
\begin{equation}
\left\vert ie_{\mathbf{k},x}^{1}-e_{\mathbf{k},z}^{1}\right\vert ^{2}=1,\ \
\left\vert ie_{\mathbf{k},x}^{2}-e_{\mathbf{k},z}^{2}\right\vert ^{2}=\frac{%
k_{y}^{2}}{k^{2}}.
\end{equation}%
In polar coordinates, with $k_{y}=k\cos \theta $%
\begin{eqnarray}
\frac{\Delta \omega }{\omega _{m}} &=&\frac{\gamma \mu _{0}M_{s}V_{s}\omega
_{m}^{2}}{4c^{3}}\int \sin \theta d\theta d\phi \left( 1+\cos ^{2}\theta
\right) \\
&=&\frac{\gamma \mu _{0}M_{s}V_{s}\omega _{m}^{2}}{3\pi c^{3}}.
\label{FreeRad:Res}
\end{eqnarray}
This result agrees with theory and experiments on mm sized spheres \cite%
{RD2A,RD2}.\textit{\ }

\section{Classical description of magnet-magnet coupling}

\label{Coup:Class} Here we formulate the non-local dissipative coupling of
the magnetization dynamics in a waveguide by the classical LLG equation. We
can incorporate the dynamic magnetic fields $\tilde{\mathbf{H}}%
_{2\rightarrow 1}^{r}$ and $\tilde{\mathbf{H}}_{1\rightarrow 2}^{r}$ between
two magnetic spheres as \cite{non_local}, 
\begin{align}
\frac{dM_{1,\alpha }}{dt}& =-\gamma \mu _{0}\varepsilon _{\alpha \beta
\delta }M_{1,\beta }H_{1,\delta }^{\mathrm{eff}}+\gamma \mu _{0}\varepsilon
_{\alpha \beta \delta }M_{1,\beta }\tilde{H}_{2\rightarrow 1,\delta }^{r} 
\notag \\
& +\frac{\alpha _{G}+\alpha _{1,\delta }^{r}}{M_{1,s}}\varepsilon _{\alpha
\beta \delta }M_{1,\beta }\frac{dM_{1,\delta }}{dt}, \\
\frac{dM_{2,\alpha }}{dt}& =-\gamma \mu _{0}\varepsilon _{\alpha \beta
\delta }M_{2,\beta }H_{2,\delta }^{\mathrm{eff}}+\gamma \mu _{0}\varepsilon
_{\alpha \beta \delta }M_{2,\beta }\tilde{H}_{1\rightarrow 2,\delta }^{r} 
\notag \\
& +\frac{\alpha _{G}+\alpha _{2,\delta }^{r}}{M_{2,s}}\varepsilon _{\alpha
\beta \delta }M_{2,\beta }\frac{dM_{2,\delta }}{dt}.
\end{align}%
The magnetic fields from Eq.~(\ref{eqn:out_of_phase}) read for $z_{1}>z_{2}$ 
\begin{align}
\tilde{H}_{2\rightarrow 1,\delta }^{r}(\mathbf{r}_{1},t)& =-i\frac{\mu
_{0}V_{s}}{v(k_{\omega })}\mathcal{H}_{k_{\omega },\delta }(\pmb{\rho}_{1})%
\mathcal{H}_{k_{\omega },\eta }^{\ast }(\pmb{\rho}_{2})M_{2,\eta }  \notag \\
& \times e^{ik_{\omega }(z_{1}-z_{2})}, \\
\tilde{H}_{1\rightarrow 2,\delta }^{r}(\mathbf{r}_{2},t)& =-i\frac{\mu
_{0}V_{s}}{v(k_{\omega })}\mathcal{H}_{-k_{\omega },\delta }(\pmb{\rho}_{2})%
\mathcal{H}_{-k_{\omega },\eta }^{\ast }(\pmb{\rho}_{1})M_{1,\eta }  \notag
\\
& \times e^{ik_{\omega }(z_{1}-z_{2})},
\end{align}%
with $k_{\omega }=\frac{1}{c}\sqrt{\omega ^{2}-c^{2}\gamma _{\lambda }^{2}}$%
. The in-phase and out-of-phase components contribute \textit{field-like}
and \textit{damping-like} torques, respectively. In high quality waveguides
we can tune them by the positions of the two magnets.

Linearizing the coupled LLG equations and neglecting the small intrinsic
Gilbert damping $\alpha _{G}$ yields (summation on $\eta =\{x,z\}$) 
\begin{align}
\omega M_{1,x}-(i\gamma \mu _{0}H_{\mathrm{eff},1}+\omega \alpha
_{1,z}^{r})M_{1,z}-{J}_{z\eta }M_{2,\eta }& =0,  \notag \\
\omega M_{1,z}+(i\gamma \mu _{0}H_{\mathrm{eff},1}+\omega \alpha
_{1,x}^{r})M_{1,x}+{J}_{x\eta }M_{2,\eta }& =0,  \notag \\
\omega M_{2,x}-(i\gamma \mu _{0}H_{\mathrm{eff},2}+\omega \alpha
_{2,z}^{r})M_{2,z}-{P}_{z\eta }M_{1,\eta }& =0,  \notag \\
\omega M_{2,z}+(i\gamma \mu _{0}H_{\mathrm{eff},2}+\omega \alpha
_{2,x}^{r})M_{2,x}+{P}_{x\eta }M_{1,\eta }& =0,  \label{eqn:coupled}
\end{align}%
where 
\begin{align}
{J}_{\delta \eta }& =\frac{\gamma \mu _{0}^{2}V_{s}M_{s}}{v(k_{\omega })}%
\mathcal{H}_{k_{\omega },\delta }(\pmb{\rho}_{1})\mathcal{H}_{k_{\omega
},\eta }^{\ast }(\pmb{\rho}_{2})e^{ik_{\omega }(z_{1}-z_{2})},  \notag \\
{P}_{\delta \eta }& =\frac{\gamma \mu _{0}^{2}V_{s}M_{s}}{v(k_{\omega })}%
\mathcal{H}_{-k_{\omega },\delta }(\pmb{\rho}_{2})\mathcal{H}_{-k_{\omega
},\eta }^{\ast }(\pmb{\rho}_{1})e^{ik_{\omega }(z_{1}-z_{2})}.
\end{align}%
In the rotating wave approximation and weak coupling, we recover the
equation for the eigenmodes, Eq.~(\ref{MagAmpTwo}), 
\begin{align}
&\left( i{J}_{zx}-i{J}_{xz}+{J}_{xx}+{J}_{zz}\right) (i{P}_{zx}-i{P}_{xz}+{P}%
_{xx}+{P}_{zz})/4  \notag \\
&+\left( \omega -\omega _{1}+i\omega _{1}\alpha _{1}^{r}\right) \left(
\omega -\omega _{2}+i\omega _{2}\alpha _{2}^{r}\right) =0,
\label{eqn:eigen_classical}
\end{align}%
where $\omega _{i}=\gamma \mu _{0}H_{\mathrm{eff},i}$ and $\alpha
_{i}^{r}=(\alpha _{i,x}^{r}+\alpha _{i,z}^{r})/2$. While equivalent, this
method becomes tedious when considering many coupled magnetic spheres.


\end{appendix}

\end{document}